\documentclass[aps,prc,superscriptaddress,twoside,twocolumn,nofootinbib,10pt,%
showpacs,floatfix]{revtex4-1}

\usepackage{amsmath,amssymb}
\usepackage[dvipdfmx]{graphicx}
\usepackage{slashed}
\usepackage{epstopdf}
\usepackage{ulem} 
\usepackage[usenames]{color}
\usepackage{subfigure}
\usepackage{float}

\newcommand{\Slash}[1]{{\ooalign{\hfil/\hfil\crcr$#1$}}}

\allowdisplaybreaks

\begin{document}

\title{Examination of $N^*(1535)$ as a probe to observe the partial
    restoration of chiral symmetry in nuclear matter}

\author{Daiki Suenaga}
\email{suenaga@hken.phys.nagoya-u.ac.jp}
\affiliation{Department of Physics,  Nagoya University, Nagoya, 464-8602, Japan}

\date{\today}

\newcommand\sect[1]{\emph{#1}---}
\begin{abstract}
We investigate modifications of mass and decay width of $N^*(1535)$ in nuclear matter in a chiral symmetric way. The nucleon and $N^*(1535)$ are introduced by a parity doublet model, and nuclear matter is constructed by one-loops of the nucleon and $N^*(1535)$. The decay width of $N^*(1535)$ is studied respecting chiral symmetry. Our calculations show that the partial width of $\Gamma_{N^*\to N\pi}$ is slightly broadened by a collisional broadening, and that of $\Gamma_{N^*\to N\eta}$ is drastically suppressed at density. As a result, the total decay width $\Gamma_{\rm tot}$ gets small at density. These modifications, especially the drastic narrowing of partial width of $\Gamma_{N^*\to N\eta}$, together with the dropping of mass of $N^*(1535)$ provide experiments for observing the partial restoration of chiral symmetry in nuclear matter by means of $N^*(1535)$ resonance with useful information.
\end{abstract}
\maketitle

\section{Introduction}
\label{sec:Intro}
Investigating chiral symmetry is one of the most important subjects in Quantum Chromodynamics (QCD), since hadron masses can be explained by a spontaneous breakdown of chiral symmetry. Although chiral symmetry is broken in the vacuum, it is expected to be restored at temperature and/or density, so that the search for chiral symmetry at such extreme environments is receiving attention recently.

One of a powerful tool to investigate relations between hadron properties and chiral symmetry is a hadron effective model. One example is the linear sigma model which was introduced by Gell-Mann and Levy~\cite{GellMann:1960np,Schwinger:1957em}. In this model, a linear representation of $SU(2)_L\times SU(2)_R$ group is employed for the nucleon, and mass generation of the nucleon is demonstrated. The linear representation was extended to the parity doublet model~\cite{Detar:1988kn}. In this model, a positive-parity nucleon and an excite negative-parity nucleon, such as the nucleon and $N^*(1535)$, can be studied collectively by employing a mirror assignment. Under this assignment, the negative-parity nucleon is regarded as a chiral partner to the positive-parity nucleon, so that these nucleons get degenerated when the chiral symmetry restoration occurs. This idea was further extended to other excited states and delta isobars~\cite{Nemoto:1998um,Jido:1998av,Jido:2001nt,Jido:1999hd}.

Some theoretical studies based on hadron effective models show a tendency of chiral restoration in nuclear matter~\cite{Cohen:1991nk,Birse:1994cz}. Other chiral effective models such as the Nambu-Jona-Lasinio (NJL) model read the same tendency (see Ref.~\cite{Hatsuda:1994pi} for a review and references therein). Besides, modifications of hadrons in nuclear matter as probes to understand the partial restoration of chiral symmetry were studied~\cite{Cabrera:2000dx,Rapp:2009yu}. Experiments to observe the partial restoration of chiral symmetry by means of light mesons also exists (see Ref.~\cite{Hayano:2008vn} for a review and references therein). Investigating the chiral symmetry in nuclear matter by focusing on modifications of charmed mesons were done as well~\cite{Suenaga:2014sga,Harada:2016uca,Suenaga:2017deu}.

In order to see an indication of chiral restoration in nuclear matter, it is worth focusing on modifications of properties of $N^*(1535)$ in nuclear matter. As is well known, the observed branching ratio of $N^*\to N\eta$ mode is $\Gamma_{N^*\to N\eta}/\Gamma_{\rm tot} \sim 50$ \% in the vacuum while threshold of $N+\eta$ is closed to the mass of $N^*(1535)$, which implies $N^*(1535)$ is strongly coupled to $\eta$ meson and the nucleon. Hence, the partial restoration of chiral symmetry in nuclear matter can have a significant influence on the decay properties of $N^*\to N\eta$, since mass difference between $N^*(1535)$ and the nucleon gets small as chiral symmetry is restored, by regarding $N^*(1535)$ as a chiral partner to the nucleon within the parity doublet model. Studies on $N^*(1535)$ paying attention to chiral symmetry in-medium exist~\cite{Ghosh:2014nba,Kim:1998upa,Ohtani:2016pyk}, and on $N^*(1535)$ in the $\eta$ photo-production on nuclei are also done~\cite{Yorita:2000bu,Maruyama:2002fz,Lehr:2003km}. In Ref.~\cite{Aarts:2015mma}, a lattice calculation for investigating masses of parity partner of nucleons at temperature is performed, and a tendency of degeneracy is observed.

Studies on $N^*(1535)$ is important in the context of $\eta$ mesic nuclei which was first reported by Haider and Liu~\cite{Haider:1986sa} as well, since $N^*(1535)$ is strongly coupled to $\eta N$ system as we have stated above, and it is expected that $N^*(1535)$ resonance plays a significant role to form a $\eta$-nucleus bound state. The quest for such exotic nucleus has been animatedly performed theoretically~\cite{Jido:2002yb,Nagahiro:2003iv,Jido:2008ng,Nagahiro:2008rj,Friedman:2013zfa} and experimentally~\cite{Chrien:1988gn,Pfeiffer:2003zd,Budzanowski:2008fr}. Such studies can also lead to understanding of the partial restoration of chiral symmetry in nuclear matter in laboratories.

In Ref.~\cite{Motohiro:2015taa}, a parity doublet model in which $\omega$ meson and $\rho$ meson contributions as well as $\sigma$ meson and pion are included was considered, and the properties of nuclear matter, i.e., the nuclear saturation density, the binding energy, the
incompressibility and the symmetry energy, can be reproduced within this model at mean field level. In this paper, we extended the parity doublet model in Ref.~\cite{Motohiro:2015taa} by taking fluctuations, and calculate modifications of mass and decay width of $N^*(1535)$ in nuclear matter, to provide useful information to observe the partial restoration of chiral symmetry in nuclear matter.

This paper is organized as following. In Sec.~\ref{sec:PDM}, we construct the parity doublet model which were proposed in Ref.~\cite{Motohiro:2015taa}, and determine model parameters. Derivative couplings and $NN^*\eta$ couplings are also included to explain the decay properties of $N^*(1535)$ in the vacuum. Besides, we solve a gap equation with respect to mean field of $\sigma$ meson in nuclear matter, and study mass modifications of the nucleon and $N^*(1535)$ at density. In Sec.~\ref{sec:InMedium}, we formulate a way to incorporate fluctuations of the light mesons respecting chiral symmetry in our model. In Sec.~\ref{sec:Results}, we demonstrate a way to calculate the decay width of $N^*(1535)$ in nuclear matter, and show results. In Sec.~\ref{sec:Conclusions}, we summarize the present study and give some discussions.

\section{Model construction}
\label{sec:PDM}

\subsection{Lagrangians}
\label{sec:Lagrangian}
In the present study, we shall investigate modifications of mass and decay properties of $N^*(1535)$ in nuclear matter. In order to treat $N^*(1535)$ and the nucleon collectively, we construct an effective model for $N^*(1535)$ and the nucleon within the parity doublet model~\cite{Detar:1988kn} in this section. In this model, a nucleon which carries positive-parity is regarded as a chiral partner of a nucleon which carries negative-parity, so that the masses of them get degenerated when chiral symmetry is restored. Here, we regard the nucleon as the positive-parity nucleon while $N^*(1535)$ as the negative-parity nucleon.

The nucleon and $N^*(1535)$ are introduced via two Fermion fields $\psi_1$ and $\psi_2$ which transform under the $SU(2)_L \times SU(2)_R$ chiral transformation as
\begin{eqnarray}
\psi_{1l} \to g_L\psi_{1l}\ &,&\ \ \psi_{1r} \to g_R\psi_{1r}\ , \nonumber\\
\psi_{2l} \to g_R\psi_{2l}\ &,&\ \ \psi_{2r} \to g_L\psi_{2r} \ .
\end{eqnarray}
$\psi_{1l}$, $\psi_{1r}$, $\psi_{2l}$ and $\psi_{2r}$ are defined by 
\begin{eqnarray}
\psi_{1l(2l)} &=&\frac{1-\gamma_5}{2}\psi_{1(2)}\ , \nonumber\\
\psi_{1r(2r)} &=&\frac{1+\gamma_5}{2}\psi_{1(2)}\ , \label{LeftRight}
\end{eqnarray}
and $g_L$ and $g_R$ are elements of $SU(2)_L$ and $SU(2)_R$, respectively. To start with, we need to construct a Lagrangian which is invariant in terms of $SU(2)_L\times SU(2)_R$ chiral symmetry, parity and charge conjugation. By introducing a chiral field $M$ including $\sigma$ meson and pion which transforms under the $SU(2)_L \times SU(2)_R$ chiral transformation as
\begin{eqnarray}
M \to g_L M g_R^\dagger\ \ ,
\end{eqnarray}
Lagrangian up to $O(\partial M^2)$ is given by
\begin{widetext}
\begin{eqnarray}
{\cal L}_N &=& \bar{\psi}_{1r}(i\Slash{\partial}+\gamma_0\mu_B-g_\omega\Slash{\omega})\psi_{1r}+\bar{\psi}_{1l}(i\Slash{\partial}+\gamma_0\mu_B-g_\omega\Slash{\omega})\psi_{1l}  \nonumber\\
&+&  \bar{\psi}_{2r}(i\Slash{\partial}+\gamma_0\mu_B-g_\omega\Slash{\omega})\psi_{2r}+\bar{\psi}_{2l}(i\Slash{\partial}+\gamma_0\mu_B-g_\omega\Slash{\omega})\psi_{2l} \nonumber\\
&-& m_0\left[\bar{\psi}_{1l}\psi_{2r}-\bar{\psi}_{1r}\psi_{2l}-\bar{\psi}_{2l}\psi_{1r}+\bar{\psi}_{2r}\psi_{1l}\right]  \nonumber\\
&-& g_1\left[\bar{\psi}_{1r}M^\dagger\psi_{1l}+\bar{\psi}_{1l}M\psi_{1r}\right] - g_2\left[\bar{\psi}_{2r}M\psi_{2l}+\bar{\psi}_{2l}M^\dagger\psi_{2r}\right] \nonumber\\ 
&-& ih_1\left[\bar{\psi}_{1l}(M\Slash{\partial}M^\dagger-\Slash{\partial}MM^\dagger)\psi_{1l} + \bar{\psi}_{1r}(M^\dagger\Slash{\partial}M-\Slash{\partial}M^\dagger M)\psi_{1r}\right] \nonumber\\
&-& ih_2\left[\bar{\psi}_{2r}(M\Slash{\partial}M^\dagger-\Slash{\partial}MM^\dagger)\psi_{2r} + \bar{\psi}_{2l}(M^\dagger\Slash{\partial}M-\Slash{\partial}M^\dagger M)\psi_{2l}\right] . \label{PD1}
\end{eqnarray}
\end{widetext}
$g_\omega$, $m_0$, $g_1$,$g_2$, $h_1$ and $h_2$ in Eq.~(\ref{PD1}) are real parameters which will be determined later, and $\mu_B$ is a baryon number chemical potential added to access to finite density. Note that $\omega$ meson is introduced as a chiral singlet in the context of $SU(2)_L\times SU(2)_R$ chiral group. 

The Lagrangian for $\sigma$ meson and pion can be given by
\begin{eqnarray}
{\cal L}_{\rm M} &=& \frac{1}{4}{\rm Tr}[\partial_\mu M\partial^\mu M^\dagger]+\frac{\bar{\mu}^2}{4}{\rm Tr}[MM^\dagger]-\frac{\lambda}{16}\left({\rm Tr}[MM^\dagger]\right)^2 \nonumber\\
&& +\frac{\lambda_6}{48}\left({\rm Tr}[MM^\dagger]\right)^3 + \frac{\epsilon}{4}\left({\rm Tr}[{\cal M}^\dagger M]+{\rm Tr}[M^\dagger{\cal M}]\right) \nonumber\\
&& -\frac{1}{4}\omega_{\mu\nu}\omega^{\mu\nu}+\frac{1}{2}m_\omega^2\omega_\mu\omega^\mu\ , \nonumber\\ \label{LMeson}
\end{eqnarray}
where $\bar{\mu}$, $\lambda$, $\lambda_6$, and $\epsilon$ are real parameters. $\omega_{\mu\nu}$ is the field strength for $\omega$ meson defined by $\omega_{\mu\nu}=\partial_\mu\omega_\nu-\partial_\nu\omega_\mu$ and $m_\omega=783$ MeV is the mass of $\omega$ meson. The second term in the second line in Eq.~(\ref{LMeson}) which explicitly breaks chiral symmetry is added to reproduce the finite mass of pion. ${\cal M}$ represents a current quark mass matrix which takes the form of
\begin{eqnarray}
{\cal M}= \left(
\begin{array}{cc}
\bar{m} & 0 \\
0 & \bar{m} \\
\end{array}
\right)\ ,
\end{eqnarray}
when the isospin symmetry is assumed.
 
In the present analysis, we follow Ref.~\cite{Motohiro:2015taa} to determine model parameters.
In this reference, the polar decomposition for the chiral field was utilized: $M=\sigma U$ with $U={\rm exp}\left(i\pi^a\tau^a/f_\pi\right)$ ($\tau^a$ is the Pauli matrix and $a$ runs over $a=1,2,3$, and $f_\pi$ is the pion decay constant), since we expect that the normal nuclear matter density is separated by the chiral restoration point. Then, by employing this procedure and separating the meson fields into their mean fields and fluctuations, we can rewrite Lagrangians~(\ref{PD1}) and~(\ref{LMeson}) as
\begin{widetext}
\begin{eqnarray}
{\cal L}_N &=& \bar{\psi}_{1r}\big(i\Slash{\partial}+\gamma_0(\mu_B-g_\omega\omega_0)\big)\psi_{1r}+\bar{\psi}_{1l}\big(i\Slash{\partial}+\gamma_0(\mu_B-g_\omega\omega_0)\big)\psi_{1l}  \nonumber\\
&+&\bar{\psi}_{2r}\big(i\Slash{\partial}+\gamma_0(\mu_B-g_\omega\omega_0)\big) \psi_{2r}+\bar{\psi}_{2l}\big(i\Slash{\partial}+\gamma_0(\mu_B-g_\omega\omega_0)\big)\psi_{2l} \nonumber\\
&-&m_0\left[\bar{\psi}_{1l}\psi_{2r}-\bar{\psi}_{1r}\psi_{2l}-\bar{\psi}_{2l}\psi_{1r}+\bar{\psi}_{2r}\psi_{1l} \right]-g_1\sigma_0\left[\bar{\psi}_{1r}\psi_{1l}+\bar{\psi}_{1l}\psi_{1r}\right]-g_2\sigma_0\left[\bar{\psi}_{2r}\psi_{2l}+\bar{\psi}_{2l}\psi_{2r}\right] \nonumber\\
&-& g_1\left[\bar{\psi}_{1r}\left(\sigma-i\frac{\sigma_0}{f_\pi}\pi\right)\psi_{1l}+\bar{\psi}_{1l}\left(\sigma+i\frac{\sigma_0}{f_\pi}\pi\right)\psi_{1r}\right]-g_2\left[\bar{\psi}_{2r}\left(\sigma+i\frac{\sigma_0}{f_\pi}\pi\right)\psi_{2l}+\bar{\psi}_{2l}\left(\sigma-i\frac{\sigma_0}{f_\pi}\pi\right)\psi_{2r}\right]  \nonumber\\
&+& \frac{g_1}{2\sigma_0}\frac{\sigma_0^2}{f_\pi^2} \left[\bar{\psi}_{1r}\pi^2\psi_{1l}+\bar{\psi}_{1l}\pi^2\psi_{1r}\right]+\frac{g_2}{2\sigma_0}\frac{\sigma_0^2}{f_\pi^2} \left[\bar{\psi}_{2r}\pi^2\psi_{2l}+\bar{\psi}_{2l}\pi^2\psi_{2r}\right] \nonumber\\
&+& h_1\left[-\frac{2\sigma_0^2}{f_\pi}\bar{\psi}_{1l}\Slash{\partial}\pi\psi_{1l}+\frac{2\sigma_0^2}{f_\pi}\bar{\psi}_{1r}\Slash{\partial}\pi\psi_{1r}-i\frac{\sigma_0^2}{f_\pi^2}\bar{\psi}_{1l}[\pi,\Slash{\partial}\pi]\psi_{1l}-i\frac{\sigma_0^2}{f_\pi^2}\bar{\psi}_{1r}[\pi,\Slash{\partial}\pi]\psi_{1r}\right] \nonumber\\
&+& h_2\left[-\frac{2\sigma_0^2}{f_\pi}\bar{\psi}_{2r}\Slash{\partial}\pi\psi_{2r}+\frac{2\sigma_0^2}{f_\pi}\bar{\psi}_{2l}\Slash{\partial}\pi\psi_{2l}-i\frac{\sigma_0^2}{f_\pi^2}\bar{\psi}_{2r}[\pi,\Slash{\partial}\pi]\psi_{2r}-i\frac{\sigma_0^2}{f_\pi^2}\bar{\psi}_{2l}[\pi,\Slash{\partial}\pi]\psi_{2l}\right]  \nonumber\\
&+& \cdots\ , \label{PolarN}
\end{eqnarray}
and
\begin{eqnarray}
{\cal L}_M &=& \frac{1}{2}\partial_\mu \sigma\partial^\mu \sigma+\frac{\sigma_0^2}{2f_\pi^2}\partial_\mu\pi^a\partial^\mu\pi^a+\frac{\bar{\mu}^2}{2}(\sigma_0+\sigma)^2-\frac{\lambda}{4}(\sigma_0+\sigma)^4+\frac{\lambda_6}{6}(\sigma_0+\sigma)^6+\bar{m}\epsilon(\sigma_0+\sigma)-\frac{\epsilon\bar{m}}{2}\frac{\sigma_0}{f_\pi^2}\pi^a\pi^a\nonumber\\
&+&\frac{1}{2}m_\omega^2\omega_0^2+ \cdots\ , \label{PolarM}
\end{eqnarray} 
\end{widetext}
respectively. In obtaining Eqs.~(\ref{PolarN}) and~(\ref{PolarM}), we have assumed a spatially homogeneous and parity non-breaking spontaneous breakdown of chiral symmetry, so that $M$ is replaced as $M= \sigma U \to (\sigma_0+\sigma)U$ in which the mean field $\sigma_0$ does not depend on space-time coordinate. $\omega_0$ is the mean field of the time-component of $\omega$ meson, and we do not include fluctuation of $\omega$ meson in the present study. 
In the vacuum, $\sigma_0\to f_\pi$ while $\omega_0\to0$.
The ellipses in Eqs.~(\ref{PolarN}) and~(\ref{PolarM}) include higher order of interactions. In order to define the mass eigenstate of positive-parity nucleon $N_+$ and negative-parity nucleon $N_-$, we need to diagonalize the mass matrix in Lagrangian~(\ref{PolarN}) by introducing a mixing angle $\theta$:
\begin{eqnarray}
\left(
\begin{array}{c}
N_+ \\
N_- \\
\end{array}
\right) = \left(
\begin{array}{cc}
{\rm cos}\, \theta & \gamma_5{\rm sin}\, \theta \\
-\gamma_5{\rm sin}\, \theta & {\rm cos}\, \theta \\
\end{array}
\right)\left(
\begin{array}{c}
\psi_1 \\
\psi_2\\
\end{array}
\right)\ . \label{Angle}
\end{eqnarray}
As already mentioned, $N_+$ is regarded as the nucleon while $N_-$ is regarded as $N^*(1535)$. Hence, using Eqs.~(\ref{LeftRight}) and~(\ref{Angle}), Lagrangians~(\ref{PolarN}) and~(\ref{PolarM}) yield
\begin{widetext}
\begin{eqnarray}
{\cal L}_N&=& \bar{N}_+\left(i\Slash{\partial}+\gamma_0\mu_B^*\right)N_++\bar{N}_-\left(i\Slash{\partial}+\gamma_0\mu_B^*\right)N_--m_+\bar{N}_+N_+-m_-\bar{N}_-N_- \nonumber\\
&-& g_{NN\sigma}\bar{N}_+\sigma N_+-g_{NN\pi}\bar{N}_+i\gamma_5\pi_r N_+
+g_{NN^*\sigma}\bar{N}_+\gamma_5\sigma N_-+g_{NN^*\pi}\bar{N}_+i\pi_r N_- \nonumber\\
&-&g_{NN^*\sigma}\bar{N}_-\gamma_5\sigma N_+-g_{NN^*\pi}\bar{N}_-i\pi_r N_+
-g_{N^*N^*\sigma}\bar{N}_-\sigma N_--g_{N^*N^*\pi}\bar{N}_-i\gamma_5\pi_r N_- \nonumber\\
&+& \frac{g_{NN\sigma}}{2\sigma_0}\bar{N}_+\pi_r^2 N_+-\frac{g_{NN^*\sigma}}{2\sigma_0}\bar{N}_+\gamma_5\pi_r^2 N_-+\frac{g_{NN^*\sigma}}{2\sigma_0}\bar{N}_-\gamma_5\pi_r^2 N_++\frac{g_{N^*N^*\sigma}}{2\sigma_0}\bar{N}_-\pi_r^2 N_- \nonumber\\
&+& 2\sigma_0h_{NN\pi}\bar{N}_+\Slash{\partial}\pi_r\gamma_5N_++2\sigma_0h_{NN^*\pi}\bar{N}_+\Slash{\partial}\pi_r N_-+2\sigma_0h_{NN^*\pi}\bar{N}_-\Slash{\partial}\pi_r N_+ + 2\sigma_0\bar{N}_-\Slash{\partial}\pi_r \gamma_5 N_- \nonumber\\
&-& if_\pi h_{NN\pi\pi}\bar{N}_+[\pi_r,\Slash{\partial}\pi_r]N_+-if_\pi h_{NN^*\pi\pi}\bar{N}_+[\pi_r,\Slash{\partial}\pi_r]\gamma_5N_- \nonumber\\
&-& if_\pi h_{NN^*\pi\pi}\bar{N}_-[\pi_r,\Slash{\partial}\pi_r]\gamma_5N_+-if_\pi h_{N^*N^*\pi\pi}\bar{N}_-[\pi_r,\Slash{\partial}\pi_r]N_- \nonumber\\
 &+& \cdots\ , \label{PDRenormN}
\end{eqnarray}
and
\begin{eqnarray}
{\cal L}_M&=&  \frac{1}{2}\partial_\mu\sigma\partial^\mu\sigma-\frac{1}{2}m_\sigma^2+\frac{1}{2}\partial_\mu\pi_r^a\partial^\mu\pi_r^a-\frac{1}{2}m_\pi^2\pi_r^a\pi_r^a \nonumber\\
&+&\frac{1}{2}\bar{\mu}^2\sigma_0^2-\frac{1}{4}\lambda\sigma_0^4+\frac{1}{6}\lambda_6\sigma_0^6+\bar{m}\epsilon\sigma_0+ \frac{1}{2}m_\omega^2\omega_0^2+\left(\bar{\mu}^2\sigma_0-\lambda\sigma_0^3+\lambda_6\sigma_0^5+\bar{m}\epsilon\right)\sigma  +\cdots\ . \nonumber\\ \label{PDRenormM}
\end{eqnarray}
\end{widetext}
In Eq.~(\ref{PDRenormN}), we have defined an effective chemical potential $\mu_B^*$ by $\mu_B^*=\mu_B-g_\omega\omega_0$. The coupling constants in Eq.~(\ref{PDRenormN}) are expressed by the mixing angle $\theta$ and original ones: $g_1$, $g_2$, $h_1$ and $h_2$ as
\begin{eqnarray}
g_{NN\sigma} &=& g_1{\rm cos}^2\theta-g_2{\rm sin}^2\theta \nonumber\\
g_{NN\pi} &=& g_1{\rm cos}^2\theta+g_2{\rm sin}^2\theta \nonumber\\
g_{NN^*\sigma} &=& g_1{\rm sin}\, \theta\, {\rm cos}\, \theta+g_2{\rm sin}\, \theta\, {\rm cos}\, \theta \nonumber\\
g_{NN^*\pi} &=& g_1{\rm sin}\, \theta\, {\rm cos}\, \theta-g_2{\rm sin}\, \theta\, {\rm cos}\, \theta \nonumber\\
g_{N^*N^*\sigma} &=& -g_1{\rm sin}^2\theta+g_2{\rm cos}^2\theta \nonumber\\
g_{N^*N^*\pi} &=& -g_1{\rm sin}^2\theta-g_2{\rm cos}^2\theta \nonumber\\
h_{NN\pi} &=& h_1{\rm cos}^2\theta-h_2{\rm sin}^2\theta \nonumber\\
h_{NN^*\pi} &=& -(h_1{\rm sin}\, \theta{\rm cos}\, \theta+h_2{\rm sin}\, \theta{\rm cos}\, \theta) \nonumber\\
h_{N^*N^*\pi} &=& h_1{\rm sin}^2\theta-h_2{\rm cos}^2\theta \nonumber\\
h_{NN\pi\pi} &=& h_1{\rm cos}^2\theta+h_2{\rm sin}^2\theta \nonumber\\
h_{NN^*\pi\pi} &=& -(h_1{\rm sin}\, \theta{\rm cos}\, \theta-h_2{\rm sin}\, \theta{\rm cos}\, \theta) \nonumber\\
h_{N^*N^*\pi\pi} &=& h_1{\rm sin}^2\theta+h_2{\rm cos}^2\theta\ ,
\end{eqnarray}
and $\theta$ satisfies a relation
\begin{eqnarray}
{\rm tan}2\theta=\frac{2m_0}{(g_1+g_2)\sigma_0}\ .
\end{eqnarray}
$m_+$ and $m_-$ in Eq~(\ref{PDRenormN}) indicate the masses of $N_+$ (the nucleon) and $N_-$ ($N^*(1535)$), respectively, and they are given by
\begin{eqnarray}
m_\pm  = \frac{1}{2}\left[\sqrt{(g_1+g_2)^2\sigma_0^2+4m_0^2}\mp(g_2-g_1)\sigma_0\right] \ . \label{NMasses} 
\end{eqnarray}
The ``bare'' $\sigma$ meson mass ($m_\sigma^2$) and pion mass ($m_\pi^2$) in Eq~(\ref{PDRenormM}) are defined by
\begin{eqnarray}
m_\pi^2 &=&\frac{\bar{m}\epsilon}{\sigma_0}\ , \label{PiMass} \nonumber\\
m_\sigma^2 &=& -\bar{\mu}^2+3\lambda \sigma_0^2-5\lambda_6 \sigma_0^4\ . \label{SigmaMass}
\end{eqnarray}
In obtaining Eqs.~(\ref{PDRenormN}) and~(\ref{PDRenormM}), we have introduced the renormalized pion field $\pi_r^a$ by $\pi^a=Z^{1/2}\pi_r^a$ with $Z=f_\pi^2/\sigma_0^2$ in such a way that kinetic term of pion field is normalized properly. Note that the one-point function (tadpole diagram) of $\sigma$ meson appears in Lagrangian~(\ref{PDRenormM}), since we have not yet solve a gap equation with respect to $\sigma_0$ to determine an appropriate ground state. This process will be done in Sec.~\ref{sec:Parameters}. 


As is well known, $N^*(1535)$ is strongly coupled with $\eta$ meson, and the partial decay width of $N^*(1535)$ is observed as
\begin{eqnarray}
\Gamma_{N^*\to N\pi} &\sim& 75\ {\rm MeV}, \label{DecayPi}\\
\Gamma_{N^*\to N\eta} &\sim& 75 \ {\rm MeV}, \label{DecayEta}
\end{eqnarray}
with the total width of $N^*(1535)$ being $\Gamma_{\rm tot}\sim150$ MeV~\cite{Olive:2016xmw}. Although the decay width of $N^*(1535)$ includes a large uncertainty, we take the same values of $\Gamma_{N^*\to N\pi}$ and $\Gamma_{N^*\to N\eta}$ as in Ref.~\cite{Jido:2002yb}. In order to take into account the large width in Eq.~(\ref{DecayEta}), we include $\eta$ meson as a chiral singlet meson into the model, namely, we include $\eta$ mesonic term and an $\eta NN^*$ coupling term are added as~\cite{Jido:2002yb} 
\begin{eqnarray}
{\cal L}_\eta &=& \frac{1}{2}\partial_\mu\eta\partial^\mu\eta-\frac{m_\eta^2}{2}\eta^2 \nonumber\\
&+&g_{ NN^*\eta}\bar{N}_-\eta N_+ + g_{ NN^*\eta}^*\bar{N}_+\eta N_-\ , \label{EtaLagrangian}
\end{eqnarray} 
with $m_\eta$ and $g_{NN^*\eta}$ being the mass of $\eta$ meson and a complex parameter, respectively.

\subsection{Parameter determination}
\label{sec:Parameters}

In this subsection, we determine the parameters. 
We fix them by properties in nuclear matter as well as those in the vacuum as performed in Ref.~\cite{Motohiro:2015taa}. We use the nucleon mass, $N^*(1535)$ mass, pion mass, $\omega$ meson mass, $\eta$ meson mass, pion decay constant, $\Gamma_{N^*\to N\pi}$, $\Gamma_{N^*\to N\eta}$, and the nucleon axial-charge $g_A$ as inputs in the vacuum. In addition, saturation density, binding energy, incompressibility are used as inputs in nuclear matter. We summarize them in Table.~\ref{tab:ParametersV} and in Table.~\ref{tab:ParametersN}. We should note that we have an additional condition of $\frac{\partial}{\partial \rho_B}(E/A-m_+^{\rm vac})|_{\rho_B^*}=0$ to reproduce the saturation behavior.
\begin{table*}[htb]
  \begin{tabular}{c|c|c|c|c|c|c|c|c} \hline\hline 
$m_+^{\rm vac}$(MeV) & $m_-^{\rm vac}$(MeV) & $m_\pi$(MeV) & $m_\omega$(MeV) & $m_\eta$(MeV) & $f_\pi$(MeV) & $\Gamma_{N^*\to N\pi}$(MeV) & $\Gamma_{N^*\to N\eta}$(MeV) &$g_A$ \\\hline
939 & 1535  & 140 & 783 & 547 & 93 &75 & 75 & 1.27 
\\ \hline \hline
  \end{tabular}
\caption{Input parameters by properties in the vacuum. In this table, $m_+^{\rm vac}$, $m_-^{\rm vac}$, $m_\pi$, $m_\omega$, $m_\eta$, $f_\pi$, $\Gamma_{N^*\to N\pi}$, $\Gamma_{N^*\to N\eta}$, $g_A$ represent the nucleon mass, $N^*(1535)$ mass, pion mass, $\omega$ meson mass, $\eta$ meson mass, decay width of $N^*\to N\pi$, decay width of $N^*\to N\eta$, nucleon axial charge, respectively.}
  \label{tab:ParametersV}
\end{table*}
\begin{table*}[htb]
  \begin{tabular}{c|c|c} \hline\hline 
$\rho_B^*$(fm$^{-3}$) & $E/A-m_+^{\rm vac}$(MeV)& K(MeV) 
\\ \hline
0.16  & -16 & 240 \\ \hline \hline
  \end{tabular}
\caption{Input parameters by properties in nuclear matter. In this table, $\rho_B^*$, $E$, $A$, $K$ represent the normal nuclear matter density, total energy of the system, mass number, incompressibility, respectively. We should note that we have additional condition of $\frac{\partial}{\partial \rho_B}(E/A-m_+^{\rm vac})|_{\rho_B^*}=0$ to reproduce the saturation nature.}
  \label{tab:ParametersN}
\end{table*}

The decay width of $\Gamma_{N^*\to NX}$ ($X=\pi,\eta$) is calculated as
\begin{eqnarray}
\Gamma_{N^*\to NX} &=&  \frac{|g_{N^*NX}|^2}{8\pi}\frac{|\vec{q}_X|}{m_-^2}\left(F(|\vec{q}_X|,\Lambda)\right)^2\nonumber\\
&& \times\left[(m_++m_-)^2-m_\pi^2\right]  \ , \label{WidthVac}
\end{eqnarray}
where $|\vec{q}_X|$ is the momentum of an emitted particle
\begin{eqnarray}
|\vec{q}_X| = \frac{\sqrt{[m_{-}^2-(m_++m_X)^2][m_{-}^2-(m_+-m_X)^2]}}{2m_{-}}\ . \nonumber\\
\end{eqnarray}
In Eq.~(\ref{WidthVac}), we have defined $g_{N^*N\pi}=\left(g_{NN^*\pi}-2\sigma_0h_{NN^*\pi}(m_+-m_-)\right)^2$. $F(|\vec{q}_\pi|,\Lambda)$ is the form factor which takes the form of 
\begin{eqnarray}
F(|\vec{q}|,\Lambda)=\frac{\Lambda^2}{|\vec{q}|^2+\Lambda^2}\ , \label{FormFactor}
\end{eqnarray}
and this is inserted in order to take a hadron size. At first glance, the insertion of  form factor in Eq.~(\ref{FormFactor}) can lead to confusions since this form factor is not Lorentz invariant. In the present analysis, however, we shall study properties of nucleons in nuclear matter at which a rest frame can be defined, and we use the decay width in Eq.~(\ref{WidthVac}) to determine parameters for studies in nuclear matter. In this respect, the expressions of form factor in Eq.~(\ref{FormFactor}) and decay width in Eq.~(\ref{WidthVac}) are understood as an adiabatic limit from finite density regime to vacuum. Therefore, as will be explained later, the value of cutoff parameter $\Lambda$ is chosen to be $\Lambda=300$ MeV~\cite{Suenaga:2017deu} which is slightly higher than the Fermi momentum of normal nuclear matter density in this paper. The axial-charge of the nucleon is taken by introducing an axial gauge field ${\cal A}_\mu$ with the gauge principle in Eq.~(\ref{PDRenormN}) and by picking up a coefficient of $\bar{N}_+\Slash{\cal A}N_+$ coupling~\footnote{Alternatively, $g_A$ can be taken by using the Goldberger-Treiman relation: 
\begin{eqnarray}
\frac{G_{\pi NN}}{m_+}= \frac{g_A}{f_\pi}\ ,
\end{eqnarray}
where $\pi NN$ coupling $G_{\pi NN}$ is obtained as $G_{\pi NN} =g_{NN\pi}+4f_\pi m_+h_{NN\pi}$ on the on-shell of the nucleon.}. The resulting coefficient reads
\begin{eqnarray}
g_A = \frac{g_{NN\pi}f_\pi+4f_\pi^2 m_+h_{NN\pi}}{m_+}\ . \label{gA}
\end{eqnarray}

From the above equations, we can determine some parameters. The remaining parameters can be fixed by properties in nuclear matter listed in Table.~\ref{tab:ParametersN}. Nuclear matter is described by one-loops of the nucleon and $N^*(1535)$ with a mean field approximation for $\sigma$ and $\omega$ meson. From Eqs.~(\ref{PDRenormN}) and~(\ref{PDRenormM}), we can calculate the effective action $\Gamma_0[\sigma_0,\omega_0]$ in terms of the mean fields $\sigma_0$ and $\omega_0$ by performing path integrals as
\begin{eqnarray}
&&\Gamma_0[\sigma_0,\omega_0] \nonumber\\
&=& -2i\sum_{i=+,-}{\rm Tr}{\rm ln}(i\Slash{\partial}-m_i+\mu_B^*\gamma_0) \nonumber\\
&+&\int d^4x \left(\frac{1}{2}m_\omega^2\omega_0^2+\frac{1}{2}\bar{\mu}^2\sigma_0^2-\frac{1}{4}\lambda\sigma_0^4+\frac{1}{6}\lambda_6\sigma_0^6+\bar{m}\epsilon\sigma_0 \right)\ . \nonumber\\ \label{EffActionMean}
\end{eqnarray}
In Eq.~(\ref{EffActionMean}), the symbol ``Tr'' stands for the trace for Dirac spinor, space-time coordinate. The effective potential $V_0[\sigma_0,\omega_0]$ is defined by $\Gamma_0[\sigma_0,\omega_0] = -V_0[\sigma_0,\omega_0]\int d^4x$ and the mean fields $\sigma_0$ and $\omega_0$ should satisfy the following stationary conditions: 
\begin{eqnarray}
\frac{\partial V_0[\sigma_0,\omega_0]}{\partial \sigma_0}=0\ ,\ \ \frac{\partial V_0[\sigma_0,\omega_0]}{\partial \omega_0}=0\ . \label{GapEq}
\end{eqnarray}
These equations determine the ground state of the present analysis and one-point functions (tadpole diagrams) of $\sigma$ meson and $\omega$ meson are eliminated in the perturbation series around this ground state. The total energy of the system ($E$) can be calculated as
\begin{widetext}
\begin{eqnarray}
E&=& \sum_{i=+,-}\int d^3x\lim_{y\to x}\langle N_i^\dagger(y_0,\vec{y})K_i(\vec{x})N_i(x_0,\vec{x})\rangle + \int d^3x\left(-\frac{1}{2}m_\omega^2\omega_0^2-\frac{1}{2}\bar{\mu}^2\sigma_0^2+\frac{1}{4}\lambda\sigma_0^4-\frac{1}{6}\lambda_6\sigma_0^6-\bar{m}\epsilon\sigma_0 \right)-E_{\rm vac}
\nonumber\\
&=& \frac{V}{4\pi^2}\sum_{i=+,-}\left\{k_{Fi}\sqrt{k_{Fi}^{2}+m_i^2}(2k_{Fi}^{2}+m_i^2)-m_i^4 {\rm ln}\left(\frac{k_{Fi}+\sqrt{k_{Fi}^{2}+m_i^2}}{m_i}\right)\right\} \nonumber\\
&& +\frac{1}{2}m_\omega^2\omega_0^2-\frac{1}{2}\bar{\mu}^2\sigma_0^2+\frac{1}{4}\lambda\sigma_0^4-\frac{1}{6}\lambda_6\sigma_0^6 - \bar{m}\epsilon\sigma_0+\left\{ \frac{1}{2}\bar{\mu}^2f_\pi^2-\frac{1}{4}\lambda f_\pi^4+\frac{1}{6}\lambda_6f_\pi^6+\bar{m}\epsilon f_\pi \right\}\ , \label{ETotal}
\end{eqnarray}
\end{widetext}
where we have defined the energy operator for one-particle state $K_i$ as $K_i=\gamma_0(-i\vec{\gamma}\cdot\vec{\partial}+m_i+g_\omega\omega_0\gamma_0)$ $(i=+,-)$. In obtaining the second line in Eq.~(\ref{ETotal}), we have utilized the in-medium propagator in the coordinate space 
\begin{eqnarray}
{\rm T}\langle N_i(x_0,\vec{x})\bar{N}_i(y_0,\vec{y}) \rangle= \int\frac{d^4k}{(2\pi)^4}\tilde{S}_{N_i}(k_0,\vec{k}){\rm e}^{-ik\cdot (x-y)}\ ,\nonumber\\
\end{eqnarray}
(${\rm T}$ is the time-ordered product operator) where $\tilde{S}_{N_i}(k_0,\vec{k})$ is of the form~\cite{TFT}
\begin{eqnarray}
\tilde{S}_{N_i}(k_0,\vec{k}) &=& (\Slash{k}+m_i)\Big[\frac{i}{k^2-m_i^2+i\epsilon} \nonumber\\
&&-2\pi\theta(k_0)\theta(k_{Fi}-|\vec{k}|)\delta(k^2-m_i^2)\Big]\ .
\end{eqnarray}
In Eq.~(\ref{ETotal}), we have subtracted the vacuum energy $E_{\rm vac}$ to measure the energy properly. $V$ is a volume of the system and we study the infinite matter such that any quantities in this paper do not depend on $V$. Also, the mass number $A$ is simply related as $A=\rho_B^* V$. $k_{F+}$ and $k_{F-}$ are the Fermi momentum for the nucleon and $N^*(1535)$ which are defined by $\mu_B^*=\sqrt{k_{F+}^2+m_+^{2}}$ and $\mu_B^*=\sqrt{k_{F-}^2+m_-^{2}}$, respectively. The baryon number density $\rho_B$ is defined by 
\begin{eqnarray}
\rho_B = \frac{2}{3\pi^2}k_{F+}^3+\frac{2}{3\pi^2}k_{F-}^3\ .
\end{eqnarray}
The incompressibility is calculated by the well-known formula 
\begin{eqnarray}
K=9\rho_B^{*2}\frac{\partial^2}{\partial\rho_B^{2}}\left(\frac{E}{A}\right)\Big|_{\rho_B^*}=9\rho_B^*\frac{\partial\mu_B}{\partial\rho_B}\Big|_{\rho_B^*} \ . \label{incompressibility}
\end{eqnarray}

From Eqs.~(\ref{ETotal}) and~(\ref{incompressibility}) together with the saturation condition $\frac{\partial}{\partial \rho_B}(E/A-m_+^{\rm vac})|_{\rho_B^*}=0$, we can fix the remaining parameters. The resultant parameters are summarized in Table.~\ref{tab:Parameters}. We should note that only $m_0$ remains as a free parameter, and we choose it as $m_0=500, 700, 900$ MeV as examples~\cite{Motohiro:2015taa}~\footnote{Eq.~(\ref{WidthVac}) is such a quadratic equation with respect to $h_{NN^*\pi}$ that we can obtain another solution. This solution leads to, however, a relatively larger value of $NN^*\pi\pi$ coupling: $h_{NN^*\pi\pi}$, which is inconsistent with our assumption of Eqs.~(\ref{DecayPi}) and~(\ref{DecayEta}). Then we have discarded this choice. }. It is notable that $k_{F-}$ is zero for any choice of $m_0$ at the normal nuclear matter density, namely, $N^* (1535)$ does not appear at normal nuclear matter because of its large mass. This it true throughout the present analysis.

\begin{table}[htb]
  \begin{tabular}{c||r|r|r} \hline
$m_0$ (MeV) & 500 & 700 & 900  \\ \hline \hline
$g_1$ & 9.03 & 7.82 & 5.97 \\
$g_2$ & 15.5 & 14.3 & 12.4 \\
$\hat{\bar{\mu}}^2$ & 73.5 & 30.8 & 1.74 \\
$\lambda$ & 139 & 58.8 & 5.00 \\
$\hat{\lambda}_6$ & 62.9 & 25.7 & 0.952  \\
$\hat{h}_1$ & 0.108 & 0.127 &  0.145 \\
$\hat{h}_2$ & 0.336  & 0.0473 & $-0.126$
\\ \hline
  \end{tabular}
\caption{Model parameters for a given value of $m_0$.  The dimensionless parameters $\hat{\bar{\mu}}^2$, $\hat{\lambda}_6$ $\hat{h}_1$ and $\hat{h}_2$ are defined by $\hat{\bar{\mu}}=\bar{\mu}^2/f_\pi^2$, $\hat{\lambda}_6=\lambda_6\cdot f_\pi^2$, $\hat{h}_1=h_1 \cdot f_\pi^2$ and $\hat{h}_2=h_2\cdot f_\pi^2$. Here, $\bar{m}\epsilon=1.56\times 10^6$ MeV$^3$, $m_\omega=783$ MeV, $m_\eta=547$ MeV, $f_\pi=92.3$ MeV and $|g_{NN^*\eta}|=2.80$.}
  \label{tab:Parameters}
\end{table}

\subsection{Partial restoration of chiral symmetry at density}
\label{sec:GapEq}
\begin{figure}[thbp]
\centering
\includegraphics*[scale=0.7]{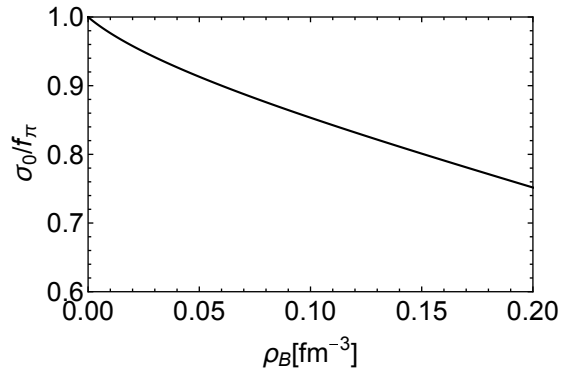}
\caption{Density dependence of the mean field $\sigma_0$ with $m_0=500$ MeV. As we can see, mean field $\sigma_0$ decreases as the density increases, which shows the tendency of partial restoration of chiral symmetry. }
\label{fig:MeanValue}
\end{figure}

Here, we show a density dependence of $\sigma_0$ and those of the masses of the nucleon and $N^*(1535)$. The density dependence of $\sigma_0$ is obtained by solving the gap equation $\partial V_0[\sigma_0,\omega_0]/\partial\sigma_0=0$ in Eq.~(\ref{GapEq}). We show it with $m_0=500$ MeV in Fig.~\ref{fig:MeanValue} as an example.
\begin{figure*}[thbp]
  \begin{center}
    \begin{tabular}{c}

  \subfigure[\ Mass of the nucleon and $N^*(1535)$]{
      \begin{minipage}{0.5\hsize}
        \begin{center}
         \includegraphics*[scale=0.55]{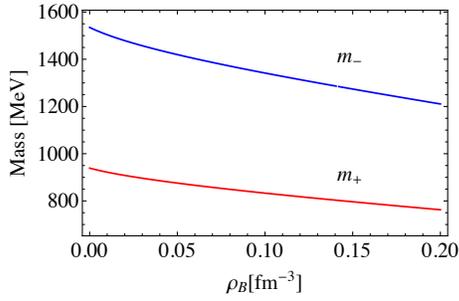}
          \hspace{0.5cm}
        \end{center}
      \end{minipage}
}
     
 \subfigure[\ Mass difference of $N^*(1535)$ and the nucleon]{
      \begin{minipage}{0.5\hsize}
        \begin{center}
          \includegraphics*[scale=0.55]{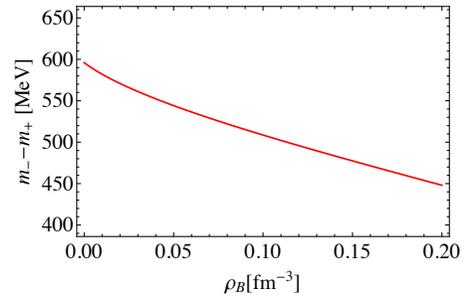}
          \hspace{0.5cm} 
        \end{center}
      \end{minipage}
}

    \end{tabular}
 \caption{(color online) Density dependences of (a) masses of the nucleon ($m_+$) (red curve) and $N^*(1535)$ ($m_-$) (blue curve), and (b) mass difference between $N^*(1535)$ and the nucleon for $m_0=500$ MeV. Mass of the nucleon decreases gradually while that of $N^*(1535)$ decreases more rapidly, so that mass difference between $N^*(1535)$ and the nucleon gets small as the density increases.}
\label{fig:NMass}
  \end{center}
\end{figure*}
As we can see, mean field $\sigma_0$ decreases as the density increases, which shows the tendency of partial restoration of chiral symmetry. Density dependences of masses of the nucleon and $N^*(1535)$, and mass difference between them, for $m_0=500$ MeV are plotted in Fig.~\ref{fig:NMass}. Red and blue curves in Fig.~\ref{fig:NMass} (a) represent masses of the nucleon and $N^*(1535)$, respectively. This figure shows that mass of the nucleon decreases gradually while that of $N^*(1535)$ decreases more rapidly, and as a result, mass difference between $N^*(1535)$ and the nucleon gets small as the density increases as shown in Fig.~\ref{fig:NMass} (b).


\section{Fluctuations of pion and $\eta$ meson at density }
\label{sec:InMedium}

\subsection{Fluctuations of pion and $\eta$ meson}
\label{sec:Fluctuation}
In Sec.~\ref{sec:PDM}, we only take the mean fields of $\sigma$ and $\omega$ meson to describe nuclear matter. Here, we extend the model to incorporate fluctuations of mesons in nuclear matter respecting chiral symmetry.

From the Lagrangians in Eqs.~(\ref{PDRenormN}),~(\ref{PDRenormM}) and~(\ref{EtaLagrangian}), the effective action around the appropriate ground state determined by the stationary conditions in Eq.~(\ref{GapEq}) is given by
\begin{eqnarray}
\Gamma [\sigma,\pi_r,\eta;\sigma_0,\omega_0] 
&=& \int D\bar{N}_+DN_+D\bar{N}_-DN_- \nonumber\\
&&\times{\rm exp}\left\{i\int d^4x\left({\cal L}_N+{\cal L}_M+{\cal L}_\eta\right)\right\} \nonumber\\
&=& \Gamma_0[\sigma_0,\omega_0] + \hat{\Gamma}[\sigma,\pi_r,\eta;\sigma_0,\omega_0]\ . \nonumber\\ \label{EffAction}
\end{eqnarray}
$\Gamma[\sigma_0,\omega_0]$ is identical to the effective action at mean field level derived in Eq.~(\ref{EffActionMean}), and $\hat{\Gamma}[\sigma,\pi_r,\eta;\sigma_0,\omega_0]$ includes the fluctuations of $\sigma$, $\pi_r$ and $\eta$ around the ground state.

\begin{figure*}[thbp]
\centering
\includegraphics*[scale=0.7]{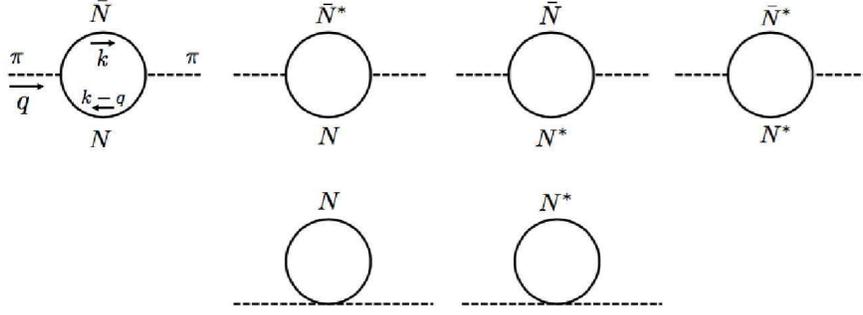}
\caption{Self-energy for pion ($\tilde{\Sigma}_\pi(q_0,\vec{q})$)}
\label{fig:Pi}
\end{figure*}
\begin{figure*}[thbp]
\centering
\includegraphics*[scale=0.5]{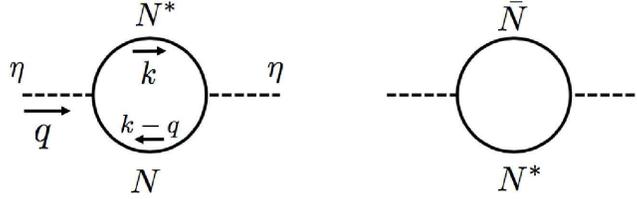}
\caption{Self-energy for $\eta$ meson ($\tilde{\Sigma}_\eta(q_0,\vec{q})$)}
\label{fig:Eta}
\end{figure*}
\begin{figure*}[thbp]
\centering
\includegraphics*[scale=0.4]{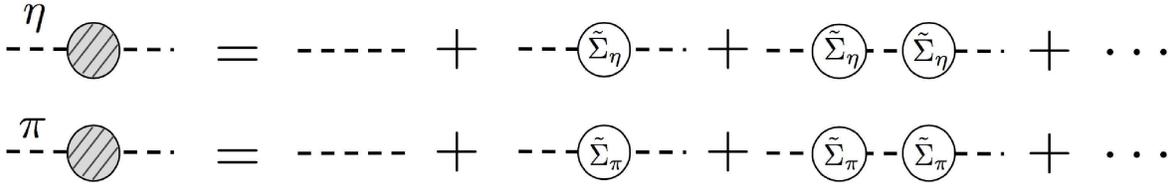}
\caption{Propagators of $\eta$ meson and pion at density. The $\tilde{\Sigma}_\eta$ and $\tilde{\Sigma}_{\pi}$ indicate the self-energies of $\eta$ meson and pion shown in Fig.~\ref{fig:Pi} and Fig.~\ref{fig:Eta}, respectively. The propagators of $\eta$ meson (pion) should include infinite sums of $\tilde{\Sigma}_{\eta}$ ($\tilde{\Sigma}_{\pi}$).}
\label{fig:ResumPropagators}
\end{figure*}


Propagators of pion and $\eta$ meson are derived by inverses of second functional derivative of the effective action in Eq.~(\ref{EffAction}) with respect to $\pi^a$ or $\eta$:
\begin{eqnarray}
\tilde{G}_\pi^{ab}(q_0,\vec{q}) &=& i\left(\int d^4x\, {\rm e}^{iq\cdot x}\frac{\delta}{\delta\pi_r^a(x)}\frac{\delta}{\delta\pi_r^b(0)}\hat{\Gamma}\right)^{-1}\nonumber\\ 
&\equiv&\frac{i\delta^{ab}}{q^2-m_\pi^2-i\tilde{\Sigma}_\pi(q_0,\vec{q})} \ , \label{TwoPi}\\ \nonumber\\ 
\tilde{G}_\eta(q_0,\vec{q}) &=& i\left(\int d^4x\, {\rm e}^{iq\cdot x}\frac{\delta}{\delta\eta(x)}\frac{\delta}{\delta\eta(0)}\hat{\Gamma}\right)^{-1}\nonumber\\
&\equiv&\frac{i}{q^2-m_\eta^2-i\tilde{\Sigma}_\eta(q_0,\vec{q})} \ ,  \label{TwoEta}
\end{eqnarray}
respectively. The ``bare'' mass $m_\pi$ in Eq.~(\ref{TwoPi}) is given by Eq.~(\ref{PiMass}), and $m_\eta$ in Eq.~(\ref{TwoEta}) is fixed as $m_\eta=547$ MeV. The self-energies $\tilde{\Sigma}_\pi(q_0,\vec{q})$ and $\tilde{\Sigma}_\eta(q_0,\vec{q})$ are diagrammatically shown in Fig.~\ref{fig:Pi} and Fig.~\ref{fig:Eta}, respectively, which are directly obtained by the formulae in Eqs.~(\ref{TwoPi}) and~(\ref{TwoEta}). The matter effects to $\eta$ meson and pion are induced by these self-energies since they include the nucleon and/or $N^*(1535)$ loops. We give the detailed expression of $\tilde{\Sigma}_\pi(q_0,\vec{q})$ and $\tilde{\Sigma}_\eta(q_0,\vec{q})$ in Appendix.~\ref{sec:SelfEnergy}. In a similar way, we can get a propagator of $\sigma$ meson on the appropriate ground state of our model. In the present analysis, however, we assume that the decay width of $N^*(1535)$ is dominated by $\Gamma_{N^*\to N\pi}$ and $\Gamma_{N^*\to N\eta}$ as in Eqs.~(\ref{DecayPi}) and~(\ref{DecayEta}). Hence, three body decay processes as $\Gamma_{N^*\to N\pi\pi}$ are neglected and $\Gamma_{N^*\to N\sigma}$ is also ignored in this sense.

We should note that the propagators of $\eta$ meson (pion) derived in Eqs.~(\ref{TwoPi}) and~(\ref{TwoEta}) claim that they should include infinite sums of the self-energies $\tilde{\Sigma}_{\eta}$ ($\tilde{\Sigma}_{\pi}$) as depicted in Fig.~\ref{fig:ResumPropagators}. In this procedure, we can fully respect chiral symmetry since the propagators are directly derived by the effective action at density in Eq.~(\ref{EffAction})~\footnote{The propagator of pion in Eq.~(\ref{TwoPi}) (diagrammatically shown in Fig.~\ref{fig:ResumPropagators}) contains a massless spectrum in the chiral limit $\epsilon\to 0$ which is regarded as the Nambu-Goldstone (NG) boson as derived in Appendix.~\ref{sec:SelfEnergy}. Therefore, we can see that the present analysis fully preserves chiral symmetry.}.

\subsection{Excitation of $\eta$ meson mode and $N^*$-hole mode in nuclear matter}
\label{sec:PHMode}
Here, we shall show appearance of another one particle state in the propagator of $\eta$ meson obtained by Eq.~(\ref{TwoEta}), which is called an $N^*$-hole mode ($N^*$-$h$ mode) in this paper in addition to the $\eta$ meson mode . The $N^*$-$h$ mode plays an important role in calculating the modification of decay width of $N^*(1535)$ as will be explained in Sec.~\ref{sec:Results}.
\begin{figure}[thbp]
\centering
\includegraphics*[scale=0.55]{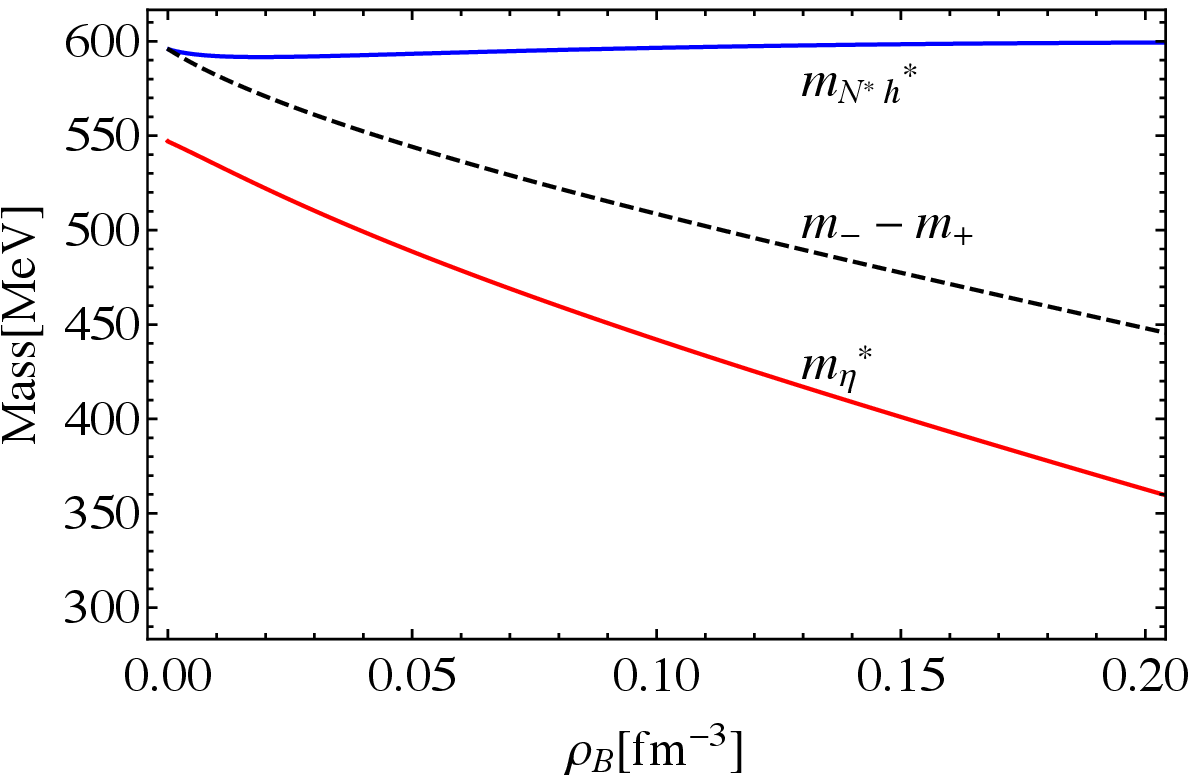}
\caption{(color online) Two solutions of Eq.~(\ref{PHMode}) with $\vec{q}=\vec{0}$. Red curve is regarded as the mass of $\eta$ meson mode since this curve reaches $q_0=547$ MeV at $\rho_B=0$ fm$^{-3}$. Blue one is identified as the mass of $N^*$-$h$ mode. Dashed curve is mass difference between $N^*(1535)$ and the nucleon, which is plotted as a reference.}
\label{fig:PHMode}
\end{figure}

One particle state of the propagator of $\eta$ meson is defined by a solution $q_0$ of the following equation:
\begin{eqnarray}
{\rm Re}\left(i\tilde{G}^{-1}_\eta(q_0,\vec{q})\right)=q^2-m_\eta^2-{\rm Re}\left(i\tilde{\Sigma}_\eta(q_0,\vec{q})\right)=0\ . \nonumber\\ \label{PHMode}
\end{eqnarray}
This equation possesses two solutions which are identified as $\eta$ meson mode and $N^*$-$h$ mode, respectively. We plot the density dependence of these solutions with $\vec{q}=\vec{0}$ which can be referred to as masses of these modes in Fig.~\ref{fig:PHMode}. Red curve is regarded as the $\eta$ meson mode since this curve reaches $q_0=547$ MeV at $\rho_B=0$ fm$^{-3}$. Blue one is identified as the $N^*$-$h$ mode. Dashed curve is mass difference between $N^*(1535)$ and the nucleon: $\Delta_{m}\equiv m_--m_+$ which is plotted as a reference. We should note that the mass of $\eta$ meson mode decreases as the density increases, and this curve always lies below $\Delta_m$. On the other hand, mass of $N^*$-$h$ mode is stable against the density, and this mode is always above $\Delta_m$.

\begin{figure}[thbp]
\centering
\includegraphics*[scale=0.55]{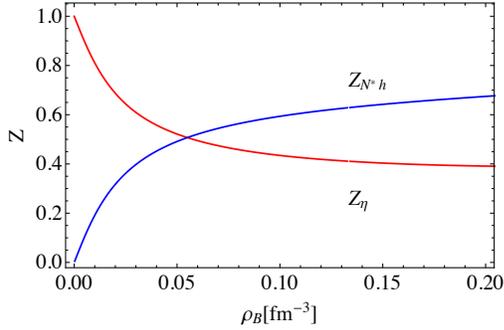}
\caption{(color online) Density dependences of $Z_\eta$ and $Z_{N^*h}$ defined by Eq.~(\ref{ZFactor}). Red curve is $Z_\eta$ ($\eta$ meson mode), and blue one is $Z_{N^*h}$ ($N^*$-$h$ mode).}
\label{fig:ZFactor}
\end{figure}
In order to study the relative strength of $\eta$ meson mode and $N^*$-$h$ mode, it is worth calculating $Z$-factors of them. They are defined by
\begin{eqnarray}
Z_{\alpha}^{-1} \equiv \frac{1}{2m_\eta}\frac{\partial}{\partial q_0}\left(q_0^2-m_\eta^2-{\rm Re}(i\tilde{\Sigma}_\eta(q_0,\vec{0}))\right)\Big|_{q_0=m^*_\alpha}\ .\nonumber\\ \label{ZFactor}
\end{eqnarray}
The subscript $\alpha$ runs over $\alpha=\eta,N^*h$, and $m_\eta^*$ stands for the mass of $\eta$ meson mode while $m_{N^*h}^*$ stands for the mass of $N^*$-$h$ mode. $\frac{1}{2m_\eta}$ in front of the derivative in Eq.~(\ref{ZFactor}) is added as a normalization. We plot density dependence of $Z_\eta$ and $Z_{N^*h}$ in Fig.~\ref{fig:ZFactor}. This figure shows that $Z_\eta$ starts from $Z_\eta=1$ at $\rho_B=0$ fm$^{-3}$, and decreases as the density increases. On the other hand, $Z_{N^*h}$ starts from $Z_{N^*h}=0$ at $\rho_B=0$ fm$^{-3}$, and increases. As a result, $Z_{N^*h}$ gets larger than $Z_{\eta}$ around $\rho_B\sim 0.055$ fm$^{-3}$, which means the strength of $N^*$-$h$ mode is stronger than that of $\eta$ meson mode at higher density. A similar tendency was reported in Ref.~\cite{Jido:2008ng}. This inversion will play a significant role in calculating the decay width of $N^*(1535)$ in Sec.~\ref{sec:Results}.

\begin{figure*}[thbp]
\centering
\includegraphics*[scale=0.6]{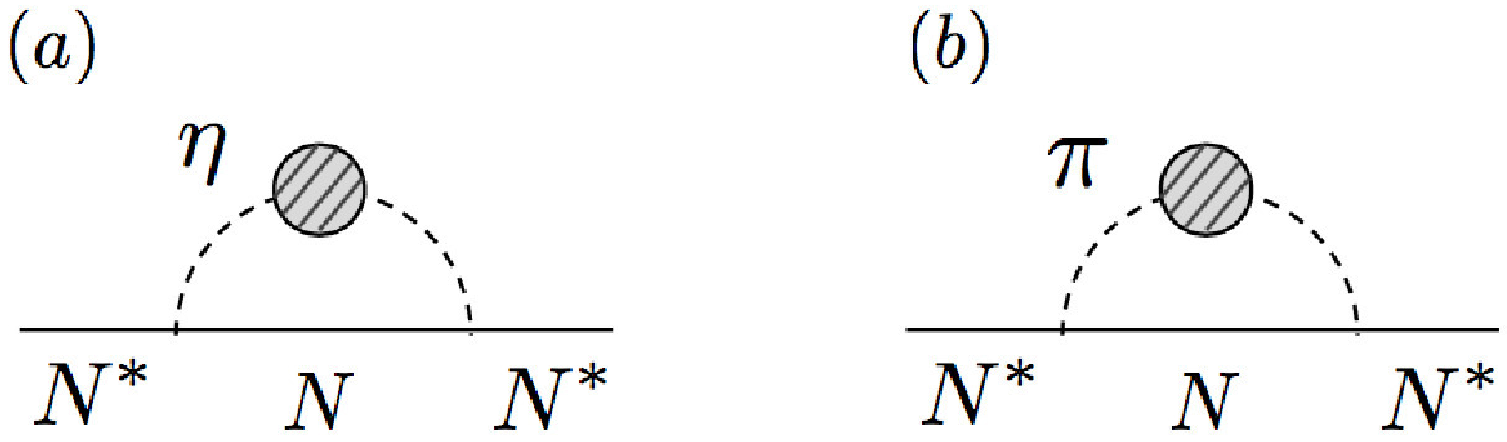}
\caption{Self-energies for $N^*(1535)$ from (a) $N\eta$ loop ($\tilde{\Sigma}_{N^*(a)}$) and (b) $N\pi$ loop ($\tilde{\Sigma}_{N^*(b)}$).  The blobs indicate the infinite sums of self-energies for $\eta$ meson ($\tilde{\Sigma}_\eta$) or pion ($\tilde{\Sigma}_\pi$) as shown in Fig.~\ref{fig:ResumPropagators}.}
\label{fig:NstSelfEnergy}
\end{figure*}
\section{Calculations and results}
\label{sec:Results}
In this section, we calculate the modification of decay width of $N^*(1535)$ in nuclear matter. In Sec.~\ref{sec:Calculations}, we show a way to calculate the decay width by computing $\Gamma_{N^*\to N\eta}$ in detail. In Sec.~\ref{sec:Width}, we show density dependence of total decay width of $N^*(1535)$ ($\Gamma_{\rm tot}$), partial decay widths of $\Gamma_{N^*\to N\pi}$ and $\Gamma_{N^*\to N\eta}$. In Sec.~\ref{sec:EtaMassDep}, we discuss on the ``bare'' $\eta$ meson mass ($m_\eta$) dependence of our results.

\subsection{Calculation method}
\label{sec:Calculations}

The partial decay widths of $N^*(1535)$ ($\Gamma_{N^*\to N\eta}$ and $\Gamma_{N^*\to N\pi}$) are given by ``imaginary parts`` of self-energies for $N^*(1535)$ by the Cutkosky rule as
\begin{eqnarray}
\Gamma_{N^*\to N\eta} &=& \frac{1}{2m_-}{\rm tr}\left[(\Slash{q}+m_+)\overline{\rm Im}\tilde{\Sigma}^R_{N^*(a)}(q_0,\vec{q})\right]\ , \label{Width} \nonumber\\
\Gamma_{N^*\to N\pi} &=& \frac{1}{2m_-}{\rm tr}\left[(\Slash{q}+m_+)\overline{\rm Im}\tilde{\Sigma}^R_{N^*(b)}(q_0,\vec{q})\right]\ , \label{WidthPi} 
\end{eqnarray}
with $q_0$ and $\vec{q}$ satisfying $\Slash{q}=m_-$. The subscript ``R'' refers to the retarded self-energy, and the (retarded) self-energies $\tilde{\Sigma}^R_{N^*(a)}$ and $\tilde{\Sigma}^R_{N^*(b)}$ are shown in Fig.~\ref{fig:NstSelfEnergy}. The blobs in this figure indicate the infinite sums of self-energies for $\eta$ meson ($\tilde{\Sigma}_\eta$) or pion ($\tilde{\Sigma}_\pi$) as shown in Fig.~\ref{fig:ResumPropagators}. In these equations, the ``imaginary part'' is defined by
\begin{eqnarray}
\overline{\rm Im} X \equiv \frac{X-\gamma_0 X^\dagger\gamma_0}{2i}\ ,
\end{eqnarray}
and ``tr'' stands for the trace for spin and isospin indices. 
Then, from Eqs.~(\ref{Width}) and~(\ref{WidthPi}), all we need to do is to get $\overline{\rm Im}\tilde{\Sigma}^R_{N^*(a)}(q_0,\vec{q})$ and $\overline{\rm Im}\tilde{\Sigma}^R_{N^*(b)}(q_0,\vec{q})$. It is possible to calculate the decay width of $N^*(1535)$ in nuclear matter by utilizing the propagators obtained in Eqs.~(\ref{TwoPi}) and~(\ref{TwoEta}). In the present study, however, we employ a more useful method so-called the ``spectral representation method''~\cite{TFT}. Here, we shall show how to utilize this method by calculating $\overline{\rm Im}\tilde{\Sigma}^R_{N^*(a)}(q_0,\vec{q})$ as an example. 

When we define a greater self-energy $\Sigma_{N^*(a)}^>(x_0,\vec{x})$ and a lesser self-energy $\Sigma_{N^*(a)}^<(x_0,\vec{x})$ through the self-energy $\Sigma_{N^*(a)}(x_0,\vec{x})$ in Fig.~\ref{fig:NstSelfEnergy} (a) as 
\begin{eqnarray}
\Sigma_{N^*(a)}(x_0,\vec{x}) = \theta(x_0)\Sigma_{N^*(a)}^>(x_0,\vec{x}) + \theta(-x_0)\Sigma_{N^*(a)}^<(x_0,\vec{x})\ , \nonumber\\ \label{GLSE}
\end{eqnarray}
and define a retarded self-energy $\Sigma_{N^*(a)}^R(x_0,\vec{x})$ by
\begin{eqnarray}
\Sigma_{N^*(a)}^R(x_0,\vec{x}) = i\theta(x_0)\left(\Sigma_{N^*(a)}^>(x_0,\vec{x})-\Sigma_{N^*(a)}^<(x_0,\vec{x})\right)\ , \nonumber\\
\end{eqnarray}
it is well known that the following relation holds~\cite{TFT}:
\begin{eqnarray}
\overline{\rm Im}\tilde{\Sigma}^R_{N^*(a)}(q_0,\vec{q}) = \frac{1}{2}\left(\tilde{\Sigma}^>_{N^*(a)}(q_0,\vec{q})-\tilde{\Sigma}^<_{N^*(a)}(q_0,\vec{q})\right)\ . \nonumber\\\label{ImRelation}
\end{eqnarray}
In Eq.~(\ref{ImRelation}), $\tilde{\Sigma}_{N^*(a)}^R(q_0,\vec{q})$, $\tilde{\Sigma}_{N^*(a)}^>(q_0,\vec{q})$ and $\tilde{\Sigma}_{N^*(a)}^<(q_0,\vec{q})$ are the Fourier transformations of $\Sigma_{N^*(a)}^R(x_0,\vec{x}) $, $\Sigma_{N^*(a)}^>(x_0,\vec{x})$ and $\Sigma_{N^*(a)}^<(x_0,\vec{x}) $, respectively.

According to Eq.~(\ref{ImRelation}), we need to find explicit forms of $\tilde{\Sigma}_{N^*(a)}^>(q_0,\vec{q})$ and $\tilde{\Sigma}_{N^*(a)}^<(q_0,\vec{q})$ to evaluate $\overline{\rm Im}\tilde{\Sigma}^R_{N^*(a)}(q_0,\vec{q})$, so that we shall give them next. When we define a greater Green's function $G_\eta^>(x_0,\vec{x})$ ($S^>_N(x_0,\vec{x})$) and a lesser Green's function $G_\eta^<(x_0,\vec{x})$ ($S^<_N(x_0,\vec{x})$) for $\eta$ meson (the nucleon) by 
\begin{eqnarray}
G_\eta(x_0,\vec{x}) &=& \theta(x_0)G_\eta^>(x_0,\vec{x}) + \theta(-x_0)G_\eta^<(x_0,\vec{x})\ , \nonumber\\
S_N(x_0,\vec{x}) &=& \theta(x_0)S^>_N(x_0,\vec{x}) +\theta(-x_0)S_N^<(x_0,\vec{x})\ , \nonumber\\
\end{eqnarray}
the self-energy for $N^*(1535)$ in Fig.~\ref{fig:NstSelfEnergy} (a) is expressed as
\begin{widetext}
\begin{eqnarray}
\Sigma_{N^*(a)}(x_0,\vec{x}) &=& (ig_{NN^*\eta})S_N(x_0,\vec{x})(ig^*_{NN^*\eta})G_{\eta}(x_0,\vec{x}) \nonumber\\
&=&  (ig_{NN^*\eta}) \left( \theta(x_0)S^>_N(x_0,\vec{x}) +\theta(-x_0)S_N^<(x_0,\vec{x})\right) (ig^*_{NN^*\eta})\left(\theta(x_0)G_\eta^>(x_0,\vec{x}) + \theta(-x_0)G_\eta^<(x_0,\vec{x})\right) \nonumber\\
&=&\theta(x_0) (ig_{NN^*\eta})S^>_N(x_0,\vec{x})(ig^*_{NN^*\eta})G_\eta^>(x_0,\vec{x}) + \theta(-x_0)(ig_{NN^*\eta})S^<_N(x_0,\vec{x})(ig^*_{NN^*\eta})G_\eta^<(x_0,\vec{x})\ . \label{SECalculation}
\end{eqnarray}
Eq.~(\ref{SECalculation}) together with Eq.~(\ref{GLSE}) reads 
\begin{eqnarray}
\Sigma_{N^*(a)}^>(x_0,\vec{x}) &=&  (ig_{NN^*\eta})S^>_N(x_0,\vec{x})(ig^*_{NN^*\eta})G_\eta^>(x_0,\vec{x}) \ ,\\
\Sigma_{N^*(a)}^<(x_0,\vec{x}) &=&  (ig_{NN^*\eta})S^<_N(x_0,\vec{x})(ig^*_{NN^*\eta})G_\eta^<(x_0,\vec{x}) \ ,
\end{eqnarray}
so that $\overline{\rm Im}\tilde{\Sigma}^R_{N^*(a)}(q_0,\vec{q})$ in Eq.~(\ref{ImRelation}) is calculated as
\begin{eqnarray}
\overline{\rm Im}\tilde{\Sigma}^R_{N^*(a)}(q_0,\vec{q}) &=& \frac{1}{2}\int\frac{d^4k}{(2\pi)^4}\left(F(\vec{k};\Lambda)\right)^2\bigg\{ (ig_{NN^*\eta})\tilde{S}^>_N(k_0,\vec{k})(ig^*_{NN^*\eta})\tilde{G}_\eta^>(q_0-k_0,\vec{q}-\vec{k}) \nonumber\\
&&  \ \ \ \ \ \ \ \ \   -(ig_{NN^*\eta})\tilde{S}^<_N(k_0,\vec{k})(ig^*_{NN^*\eta})\tilde{G}_\eta^<(q_0-k_0,\vec{q}-\vec{k}) \bigg\}\ . \label{ImComp}
\end{eqnarray}
\end{widetext}
In obtaining Eq.~(\ref{ImComp}), we have inserted the form factor $F(\vec{k};\Lambda)$ defined by Eq.~(\ref{FormFactor}) to take into account the hadron size. The value of cutoff parameter $\Lambda$ is chosen to be $\Lambda=300$ MeV in the present analysis, which is slightly higher than that of Fermi momentum at normal nuclear matter density~\cite{Suenaga:2017deu}.

Furthermore, the Fourier transformations of the greater Green's function $\tilde{G}^>_\eta(q_0,\vec{q})$ and the lesser Green's function $\tilde{G}^<_\eta(q_0,\vec{q})$ are related to the spectral function for $\eta$ meson $\rho_\eta(q_0,\vec{q})$ in an equilibrium system by following relations~\cite{TFT}:
\begin{eqnarray}
\tilde{G}_\eta^>(q_0,\vec{q}) &=& (1+f(q_0))\rho_\eta(q_0,\vec{q})  \label{Bose1}\\
\tilde{G}_\eta^<(q_0,\vec{q}) &=& f(q_0)\rho_\eta(q_0,\vec{q})\ , \label{Bose2}
\end{eqnarray}
where $f(q_0)$ is the Bose-Einstein distribution function.
In a similar way, $\tilde{S}_N^>(q_0,\vec{q})$ and $\tilde{S}_N^<(q_0,\vec{q})$ are related to spectral function for the nucleon $\rho_N(q_0,\vec{q})$ as
\begin{eqnarray}
\tilde{S}_N^>(q_0,\vec{q}) &=& (1-\tilde{f}(q_0-\mu_B^*))\rho_N(q_0,\vec{q})  \label{Fermi1}\\
\tilde{S}_N^<(q_0,\vec{q}) &=& -\tilde{f}(q_0-\mu_B^*)\rho_N(q_0,\vec{q})\ ,\label{Fermi2}
\end{eqnarray}
where $\tilde{f}(q_0-\mu_B^*)$ is the Fermi-Dirac distribution function. Note that minus signs in Eqs.~(\ref{Fermi1}) and~(\ref{Fermi2}) reflect the Pauli blocking of Fermions, and we have not given the baryon chemical potential to distribution function for $\eta$ meson in Eqs.~(\ref{Bose1}) and~(\ref{Bose2}) since $\eta$ meson does not have a baryon number. In the present study, $\rho_\eta(q_0,\vec{q})$ is obtained by 
\begin{eqnarray}
&&\rho_{\eta}(q_0,\vec{q}) \nonumber\\
&=& \frac{-2\epsilon(q_0){\rm Im}\left(i\tilde{\Sigma}_{\eta}(q_0,\vec{q})\right)}{\left[q^2-m_{\pi(\eta)}^2-{\rm Re}\left(i\tilde{\Sigma}_{\eta}(q_0,\vec{q})\right)\right]^2+\left[{\rm Im}\left(i\tilde{\Sigma}_{\eta}(q_0,\vec{q})\right)\right]^2}\ , \nonumber\\ \label{RhoPiEta}
\end{eqnarray}
and $\rho_N(q_0,\vec{q})$ is of the form 
\begin{eqnarray}
\rho_N(q_0,\vec{q}) = 2\pi(\Slash{q}+m_+)\epsilon(q_0)\delta(q^2-m_+^2)\ . \nonumber\\
\end{eqnarray}
$\epsilon(q_0)$ in Eq.~(\ref{RhoPiEta}) is defined by satisfying $\epsilon(q_0)=+1(-1)$ for $q_0>0$ ($q_0<0$). Bose-Einstein distribution function $f(q_0)$ and Fermi-Dirac distribution function $\tilde{f}(q_0-\mu_B^*)$ at zero temperature take the form of
\begin{eqnarray}
f(q_0) &=& \frac{1}{{\rm e}^{q_0/T}-1}  \overset{T\to 0}{=} -\theta(-q_0) \ ,\\
\tilde{f}(q_0-\mu_B^*) &=& \frac{1}{{\rm e}^{(q_0-\mu_B^*)/T}+1} \overset{T\to0}{=} \theta(\mu_B^*-q_0)\ . \nonumber\\ \label{FermiT0}
\end{eqnarray}

Utilizing Eqs.~(\ref{Bose1}) -~(\ref{FermiT0}), $\overline{\rm Im}\tilde{\Sigma}^R_{N^*(a)}(q_0,\vec{q})$ in Eq.~(\ref{ImComp}) is finally evaluated as
\begin{widetext}
\begin{eqnarray}
\overline{\rm Im}\tilde{\Sigma}^R_{N^*(a)}(q_0,\vec{q}) &=& \frac{1}{2}\int\frac{d^4k}{(2\pi)^3} \left(F(\vec{k};\Lambda)\right)^2\left(\theta(q_0-k_0)-\theta(\mu_B^*-k_0)\right) \nonumber\\
&& \times (ig_{NN^*\eta})(\Slash{k}+m_+)\epsilon(k_0)\delta(k^2-m_+^2)(ig^*_{NN^*\eta})\rho_\eta(q_0-k_0,\vec{q}-\vec{k}) \ .
\end{eqnarray}
\end{widetext}
Then, the calculation of partial decay width of $\Gamma_{N^*\to N\eta}$ in Eq.~(\ref{Width}) is completed. In a similar way, the partial decay width of $\Gamma_{N^*\to N\pi}$ in Eq.~(\ref{WidthPi}) and total width $\Gamma_{\rm tot}$ are also obtained. 
\begin{figure*}[thbp]
  \begin{center}
    \begin{tabular}{c}

  \subfigure[\ $m_0=500$ MeV]{
      \begin{minipage}{0.32\hsize}
        \begin{center}
         \includegraphics*[scale=0.55]{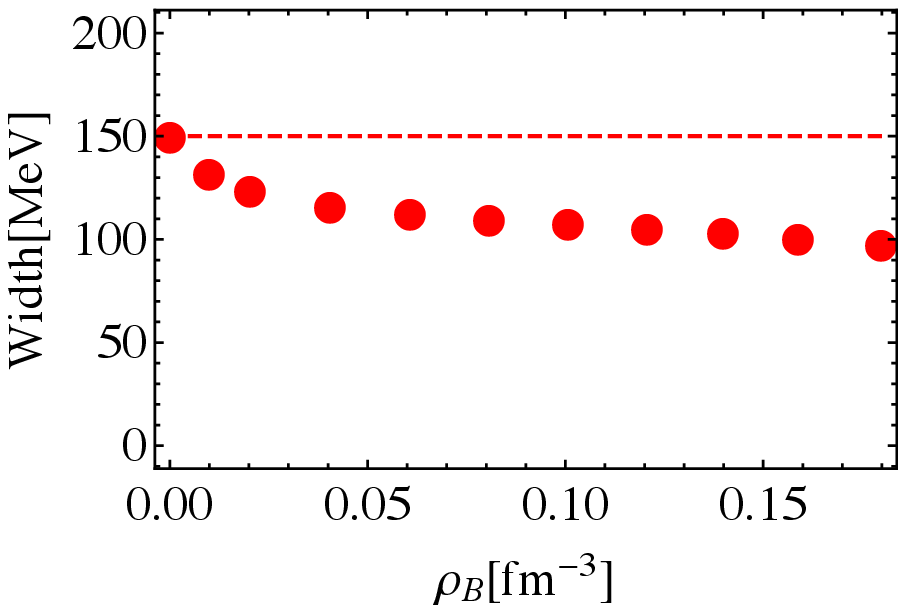}
          \hspace{0.5cm}
        \end{center}
      \end{minipage}
}
     
 \subfigure[\ $m_0=700$ MeV]{
      \begin{minipage}{0.32\hsize}
        \begin{center}
          \includegraphics*[scale=0.55]{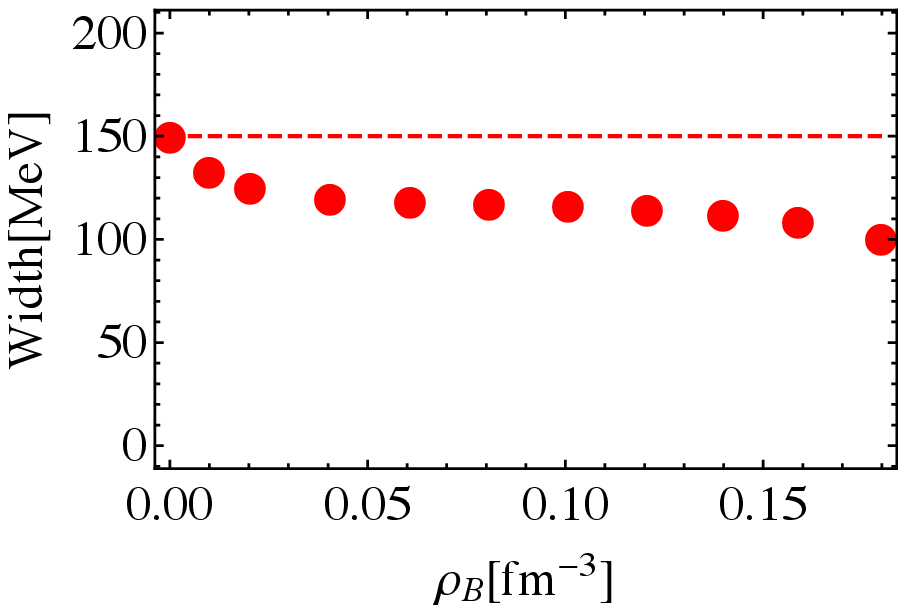}
          \hspace{0.5cm} 
        \end{center}
      \end{minipage}
}

\subfigure[\ $m_0=900$ MeV]{
      \begin{minipage}{0.32\hsize}
        \begin{center}
         \includegraphics*[scale=0.55]{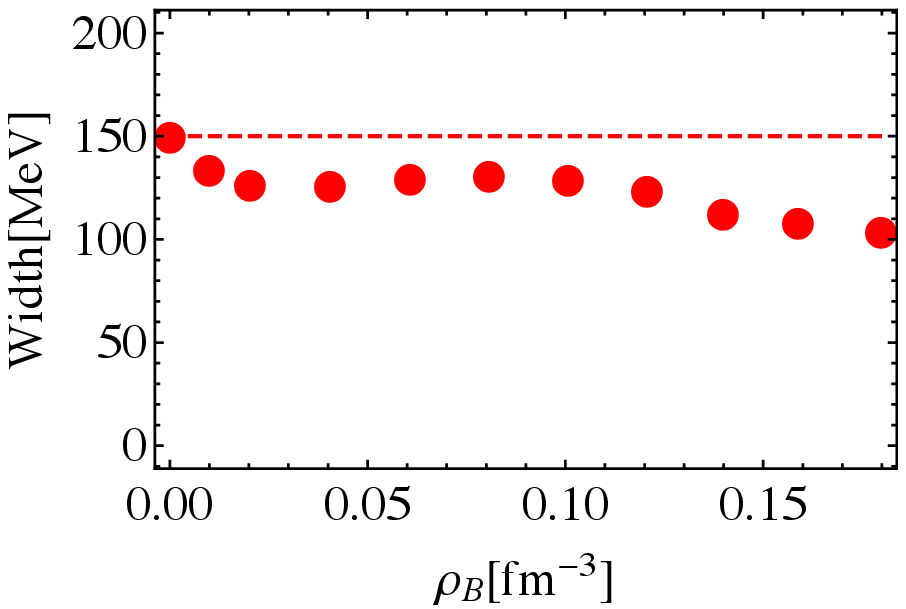}
          \hspace{0.5cm}
        \end{center}
      \end{minipage}
}

    \end{tabular}
 \caption{(color online) Density dependences of total width of $N^*(1535)$ ($\Gamma_{\rm tot}$), for (a) $m_0=500$ MeV, (b) $m_0=700$ MeV, (c) $m_0=900$ MeV. Red circles are the results, and dashed red line is the total width in the vacuum $\Gamma_{\rm tot}^{\rm vac}=150$ MeV which is added as a reference.}
\label{fig:Width}
  \end{center}
\end{figure*}
\begin{figure}[thbp]
\centering
\includegraphics*[scale=0.6]{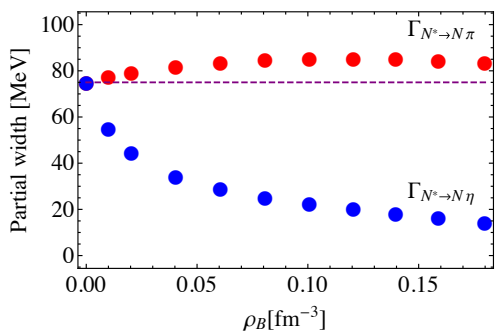}
\caption{(color online) Density dependences of partial width of $N^*(1535)$ for $m_0=500$ MeV. Red circles represent the partial width of $\Gamma_{N^*\to N\pi}$ and blue circles are $\Gamma_{N^*\to N\eta}$. Purple dashed line is the vacuum values of them: $\Gamma^{\rm vac}_{N^*\to N\pi}=\Gamma^{\rm vac}_{N^*\to N\eta}=75$ MeV, which is shown as a reference. }
\label{fig:PWidth}
\end{figure}

\subsection{Results}
\label{sec:Width}

The resultant density dependence of the total decay width of $N^*(1535)$ ($\Gamma_{\rm tot}$) in nuclear matter for $m_0=500$ MeV, $m_0=700$ MeV and $m_0=900$ MeV are plotted in Fig.~\ref{fig:Width}. Red circles are the results, and dashed red line is the total width in the vacuum $\Gamma_{\rm tot}^{\rm vac}=150$ MeV which is added as a reference. For any choice of the value of $m_0$, total decay width of $N^*(1535)$ gets small at density, and drops to about $100$ MeV at normal nuclear matter density $\rho_0=0.16$ fm$^{-3}$. This tendency shows that decay width of $N^*(1535)$ gets closed in nuclear matter although a naive expectation leads to a broadening of decay width due to collisions with nucleons surrounding $N^*(1535)$.

The origin of this narrowing can be understood well if we investigate the density dependence of partial width of $\Gamma_{N^*\to N\pi}$ and $\Gamma_{N^*\to N\eta}$. These partial widths for $m_0=500$ MeV are indicated in Fig.~\ref{fig:PWidth}. Red circles represent the partial width of $\Gamma_{N^*\to N\pi}$ and blue circles are $\Gamma_{N^*\to N\eta}$. Purple dashed line is the vacuum values of them: $\Gamma^{\rm vac}_{N^*\to N\pi}=\Gamma^{\rm vac}_{N^*\to N\eta}=75$ MeV, which is added as a reference. As we can see, the partial width of $\Gamma_{N^*\to N\pi}$ is broadened as we expect while that of $\Gamma_{N^*\to N\eta}$ is drastically closed as the density increases. This unexpected behavior of $\Gamma_{N^*\to N\eta}$ is explained as following. Although a decay process of $N^* \to N + (\eta$ meson mode) in nuclear matter is allowed since the phase space is not closed as we can see in Fig.~\ref{fig:PHMode} ($m_\eta^*$ is always below $\Delta_m=m_--m_+$), the $Z$-factor for $\eta$ meson mode ($Z_\eta$) is converted into that of $N^*$-$h$ mode ($Z_{N^*h}$) as shown in Fig.~\ref{fig:ZFactor}. The mass of $N^*$-$h$ mode is always above the mass difference $\Delta_m=m_--m_+$, so that imaginary part from decay process of $N^* \to N + (N^*$-$h$ mode) is not generated. Therefore, the main part of imaginary part in Fig.~\ref{fig:NstSelfEnergy} (a) is lost and the resulting partial width is suppressed as shown in Fig.~\ref{fig:PWidth}. 

These modifications of decay properties, especially the drastic narrowing of the partial width of $\Gamma_{N^*\to N\eta}$, together with the dropping of mass of $N^*(1535)$ provide experiments for observing the chiral restoration in nuclear matter by means of $N^*(1535)$ resonance with useful information.

\subsection{$\eta$ meson mass dependence}
\label{sec:EtaMassDep}

In the present analysis, we employ a two-flavor parity doublet model and include $\eta$ meson as a chiral singlet meson as done in Ref.~\cite{Jido:2002yb}, so that we assume the ``bare'' mass of $\eta$ meson does not change from the vacuum value $m_\eta=547$ MeV. When we extend our analysis to the three-flavor case, we expect the value of $m_\eta$ can be changed from $m_\eta=547$ MeV. Then, in this subsection, we discuss on the ``bare'' $\eta$ meson mass dependence of our results at normal nuclear matter density $\rho_0=0.16$ fm$^{-3}$ with $m_0=500$ MeV.

We show the resultant $\Gamma_{N^*\to N\eta}$ and $\Gamma_{\rm tot}$ with $m_\eta = 450$, $500$, $547$, $600$, $650$ MeV in Table.~\ref{tab:EtaMassDep}.
\begin{table}[htb]
  \begin{tabular}{c||c|c|c|c|c} \hline\hline 
$m_\eta$ (MeV)& 450 & 500& 547 & 600 & 650  \\\hline
$\Gamma_{N^*\to N\eta}$ (MeV) & 27.6  & 22.6 & 16.4 & 0 & 0 \\ \hline 
$\Gamma_{\rm tot}$ (MeV) & 112 & 107 & 101 & 84.3 & 84.3
\\ \hline \hline
  \end{tabular}
\caption{$\eta$ meson mass dependence of $\Gamma_{N^*\to N\eta}$ and $\Gamma_{\rm tot}$ at $\rho_0=0.16$ fm$^{-3}$ with $m_0=500$ MeV.}
  \label{tab:EtaMassDep}
\end{table}
When we take $m_\eta=600$ MeV and $m_\eta=650$ MeV, the decay width of $\Gamma_{N^*\to N\eta}$ is completely closed. On the other hand, when we take $m_\eta=450$ MeV and $m_\eta=500$ MeV, $\Gamma_{N^*\to N\eta}$ is slightly larger than the result with $m_\eta=547$ MeV. From this table, we can expect the result does not differ from the present analysis so much when we study the width of $N^*(1535)$ in a three-flavor model.

In the three-flavor model, couplings of $\eta NN$ and $\eta N^*N^*$ can appear and they can broaden the partial decay width of $\Gamma_{N^*\to N\eta}$. Even when these effects are included, the structure of $\eta$ meson mode and $N^*$-$h$ mode at density is qualitatively universal and we expect these corrections provide minor changes for our results.

\section{Conclusions}
\label{sec:Conclusions}
In the present study, we investigate the mass and decay width of $N^*(1535)$ in nuclear matter to give some clues to understand the partial restoration of chiral symmetry in-medium. The nucleon and $N^*(1535)$ are introduced within the parity doublet model, so that the nucleon and $N^*(1535)$ is regarded as the chiral partner to each other. Then, mass difference between $N^*(1535)$ and the nucleon is expected to get small as the density increases in which chiral restoration is realized. 

In this study, we determine model parameters to reproduce the properties of normal nuclear matter in addition to the vacuum ones as listed in Table.~\ref{tab:ParametersV} and Table.~\ref{tab:ParametersN} as done in Ref.~\cite{Motohiro:2015taa}. We also include $\eta$ meson as a chiral singlet meson, and add $NN^*\eta$ coupling to explain a large decay width of $\Gamma_{N^*\to N\eta}$ in the vacuum as in Ref.~\cite{Jido:2002yb}.

In Fig.~\ref{fig:MeanValue}, we plot density dependence of mean field of $\sigma$ meson with chiral invariant mass $m_0$ being $500$ MeV. As the density increases, the value of mean field decreases which shows a tendency of partial restoration of chiral symmetry in nuclear matter. Accordingly, mass of the nucleon decreases gradually while that of $N^*(1535)$ decreases more rapidly so that mass difference between $N^*(1535)$ and the nucleon drops as the density increases, as we can see in Fig.~\ref{fig:NMass}.

The decay width of $N^*(1535)$ in nuclear matter is studied by calculating the ``imaginary part'' of self-energy for $N^*(1535)$ in Fig.~\ref{fig:NstSelfEnergy} by the Cutkosky rule. In this figure, propagators of pion and $\eta$ meson should be ones which include infinite sums of self-energies in Fig.~\ref{fig:Pi} and Fig.~\ref{fig:Eta} to maintain the chiral symmetry in our calculations. The calculations show the width of $N^*(1535)$ is suppressed at density as shown in Fig.~\ref{fig:Width}, in contrast to the naive expectation in which collisional broadening provides an enlargement of width of $N^*(1535)$. This behavior is caused by a drastic suppression of partial decay width of $\Gamma_{N^*\to N\eta}$ as we can see in Fig.~\ref{fig:PWidth}. Although the decay process of $N^*\to N + (\eta$ meson mode) is allowed since phase space is not closed, the $Z$-factor for $\eta$ meson is converted into that of $
N^*$-$h$ mode as we access to finite density as shown in Fig.~\ref{fig:PHMode} and Fig.~\ref{fig:ZFactor}. The decay process of $N^* \to N+(N^*$-$h$ mode) is always forbidden since its phase space is not opened at any densities. Therefore, main part of imaginary part in Fig.~\ref{fig:NstSelfEnergy} (a) is lost and resulting partial width is drastically suppressed. These modifications of decay properties of $N^*(1535)$, especially the drastic narrowing of the partial width of $\Gamma_{N^*\to N\eta}$, together with the dropping of mass of $N^*(1535)$ provide experiments for observing the chiral restoration in nuclear matter by means of $N^*(1535)$ resonance with useful information.

In the following, we give some discussions which cannot be covered in this paper. In the present study, we calculate the decay width of $N^*(1535)$ in nuclear matter. In the experiment, $N^*(1535)$ is produced as a resonance state so that information of off-shell state is significant as well. From this point of view, it is interesting to study the spectral function of $N^*(1535)$ in nuclear matter, and we leave this in future publication.

In this study, we utilize one-loop approximation for describing nuclear matter and investigate the density dependence of mass and decay width of $N^*(1535)$. We estimate higher order loop-corrections to our results by the same method employed in Ref.~\cite{Suenaga:2017deu} so as to confirm the validity of our results. In terms of the mean field of $\sigma$ meson, we can see that the higher order loop-corrections are at most several MeV while the value of $\sigma_0$ is $73$ MeV with $m_0=500$ MeV and $\rho_B=0.16$ fm$^{-3}$. We also study higher order loop-corrections to the mass of $N^*(1535)$, and the result also shows the corrections are small in comparison with the one-loop level. These estimations enable us to expect the higher order loop-corrections do not provide significant changes to our results.

In this study, we employ the two-flavor parity doublet model and $\eta$ is regarded as a chiral singlet meson since chiral symmetry is explicitly broken for strange sector due to a large mass of strange quark. Then, we assume the ``bare'' mass of $\eta$ meson is not changed from $m_\eta=547$ MeV at density. When we extend our model into a three-flavor model, this value can be changed. In order to estimate this change, we study $m_\eta$ dependence of $\Gamma_{N^*\to N\eta}$ and $\Gamma_{\rm tot}$ in Sec.~\ref{sec:EtaMassDep}. As a result, we confirm that the result does not change so much. Even when the $\eta NN$ and $\eta N^*N^*$ couplings are included explicitly, the structure of $\eta$ meson mode and $N^*$-$h$ mode at density is qualitatively universal and we expect these corrections also provide minor changes for our results.

In the present study, we do not include effects of $\Delta$ while $\Delta$ couples to pion and the nucleon. We expect, however, that $\Delta$ does not play a major role in the decay width of $N^*(1535)$ since a direct $N^*(1535)\Delta\pi$ coupling is negligible~\cite{Olive:2016xmw}. Therefore, $\Delta$ has a small influence upon the decay width of $N^*(1535)$ and we expect our conclusion is not changed since our results are mainly provided by the structure of $\eta$ meson mode and $N^*$-$h$ mode.
 
 Furthermore, we assume that three body decay of $N^*(1535)$ vanishes and we ignore $\sigma$ meson decay processes. We have confirmed that such $\sigma$ meson decay provides the decay width of $N^*(1535)$ with corrections of only a few MeV at most.

\acknowledgments

This work is supported partly by the Grant-in-Aid for JSPS Research Fellow No. 17J05638. The auther would like to thank Y. Takeda for useful discussions on parity doublet model, and M. Harada for useful comments.


\appendix

\section{Self-energies for $\sigma$ meson and pion}
\label{sec:SelfEnergy}
Here, we give explicit calculations of self-energies for pion and $\eta$ meson in Fig.~\ref{fig:Pi} and Fig.~\ref{fig:Eta}. In calculating these diagrams, we employ the in-medium propagators for the nucleon and $N^*(1535)$:
\begin{widetext}
\begin{eqnarray}
\tilde{S}_{N_+}(k_0,\vec{k}) &=& (\Slash{k}+m_+) \bigg[\frac{i}{k^2-m_+^2+i\epsilon}-2\pi\theta(k_0)\theta(k_{F+}-|\vec{k}|)\delta(k^2-m_+^2)\bigg] \ ,  \\
\tilde{S}_{N_-}(k_0,\vec{k}) &=& (\Slash{k}+m_-) \bigg[\frac{i}{k^2-m_-^2+i\epsilon}  -2\pi\theta(k_0)\theta(k_{F-}-|\vec{k}|)\delta(k^2-m_-^2)\bigg] \ . \nonumber\\
\end{eqnarray} 
The resulting retarded self-energy for pion in Fig.~\ref{fig:Pi} is obtained as
\begin{eqnarray}
{\rm Re}\tilde{\Sigma}^R_{\pi} (q_0,\vec{q})&=& \frac{g_{NN\pi}^2}{2\pi^2}\int_0^{k_{F+}}d|\vec{k}|\frac{|\vec{k}|^2}{E_k^+}\left\{4-\frac{q_0^2-|\vec{q}|^2}{2|\vec{k}||\vec{q}|}{\rm ln}\left|A_+\right|\right\} - m_+^2\frac{\left(2\sigma_0h_{NN\pi}\right)^2}{\pi^2}\int_0^{k_{F+}} d|\vec{k}| \frac{|\vec{k}|^2}{E^+_k} \frac{q_0^2-|\vec{q}|^2}{|\vec{k}||\vec{q}|}{\rm ln}\left|A_+\right| \nonumber\\ \nonumber\\
&+& \frac{4\sigma_0m_+g_{NN\pi}h_{NN\pi}}{\pi^2}(q_0^2-|\vec{q}|^2)\int_0^{k_{F+}} d|\vec{k}|\frac{|\vec{k}|^2}{E_k^+}\frac{1}{2|\vec{k}||\vec{q}|}{\rm ln}\left|A_+\right| \nonumber\\ \nonumber\\
&+& \frac{g_{NN^*\pi}^2}{\pi^2}\int_0^{k_{F+}} d|\vec{k}| \frac{|\vec{k}|^2}{E_k^+}\left\{2-\frac{q_0^2-|\vec{q}|^2-(m_++m_-)^2}{4|\vec{k}||\vec{q}|}{\rm ln}\left|A_{+-}\right|\right\} \nonumber\\ \nonumber\\
&+& \frac{g_{NN^*\pi}^2}{\pi^2}\int_0^{k_{F-}} d|\vec{k}|\frac{|\vec{k}|^2}{E_k^-}\left\{2-\frac{q_0^2-|\vec{q}|^2-(m_++m_-)^2}{4|\vec{k}||\vec{q}|}{\rm ln}\left|A_{-+}\right|\right\}\nonumber\\ \nonumber\\
&+& \frac{2(2\sigma_0h_{NN^*\pi})^2}{\pi^2}\int_0^{k_{F+}} d|\vec{k}| \frac{|\vec{k}|^2}{2E_k^+} \bigg\{2(m_-^2-m_+^2) - \frac{(m_--m_+)^2(q_0^2-|\vec{q}|^2-(m_++m_-)^2)}{4|\vec{k}||\vec{q}|} {\rm ln}\left|A_{+-}\right| \biggr\} \nonumber\\ \nonumber\\
&+& \frac{2(2\sigma_0h_{NN^*\pi})^2}{\pi^2}\int_0^{k_{F-}} d|\vec{k}| \frac{|\vec{k}|^2}{2E_k^-} \bigg\{2(m_-^2-m_+^2) -\frac{(m_--m_+)^2(q_0^2-|\vec{q}|^2-(m_++m_-)^2)}{4|\vec{k}||\vec{q}|} {\rm ln}\left|A_{-+}\right| \biggr\} \nonumber\\ \nonumber\\
&+&\frac{8\sigma_0g_{NN^*\pi}h_{NN^*\pi}}{\pi^2}\int_0^{k_{F+}} d|\vec{k}|\frac{|\vec{k}|^2}{2E_{k}^+}\left\{2(m_++m_-)-\frac{(m_--m_+)(q_0^2-|\vec{q}|^2-(m_++m_-)^2)}{4|\vec{k}||\vec{q}|}\right\}  {\rm ln}\left|A_{+-}\right| \nonumber\\
&-&\frac{8\sigma_0g_{NN^*\pi}h_{NN^*\pi}}{\pi^2}\int_0^{k_{F-}} d|\vec{k}| \frac{|\vec{k}|^2}{2E_{k}^-}\left\{2(m_++m_-)+\frac{(m_--m_+)(q_0^2-|\vec{q}|^2-(m_++m_-)^2)}{4|\vec{k}||\vec{q}|}\right\} {\rm ln}\left|A_{-+}\right|\nonumber\\ \nonumber\\
 &+&\frac{g_{N^*N^*\pi}^2}{2\pi^2}\int_0^{k_{F-}}d|\vec{k}|\frac{|\vec{k}|^2}{E_k^-}\left\{4-\frac{q_0^2-|\vec{q}|^2}{2|\vec{k}||\vec{q}|}{\rm ln}\left|A_-\right|\right\} - m_-^2\frac{\left(2\sigma_0h_{N^*N^*\pi}\right)^2}{\pi^2}\int_0^{k_{F-}} d|\vec{k}| \frac{|\vec{k}|^2}{E^-_k} \frac{q_0^2-|\vec{q}|^2}{|\vec{k}||\vec{q}|}{\rm ln}\left|A_-\right| \nonumber\\ \nonumber\\
&+& \frac{4\sigma_0m_-g_{N^*N^*\pi}h_{N^*N^*\pi}}{\pi^2}(q_0^2-|\vec{q}|^2)\int_0^{k_{F-}} d|\vec{k}|\frac{|\vec{k}|^2}{E^-_k}\frac{1}{2|\vec{k}||\vec{q}|} {\rm ln}\left|A_{-}\right| \nonumber\\ \nonumber\\
&-& \frac{2g_{NN\sigma}m_+}{\pi^2\sigma_0}\int_0^{k_{F+}}d|\vec{k}|\frac{|\vec{k}|^2}{E_k^+} -\frac{2g_{N^*N^*\sigma}m_-}{\pi^2\sigma_0}\int_0^{k_{F-}}d|\vec{k}|\frac{|\vec{k}|^2}{E_k^-}\ , \label{PiReal} 
 \end{eqnarray}
and
\begin{eqnarray}
{\rm Im}\tilde{\Sigma}_{\pi}^R(q_0,\vec{q}) &=& \pi G_1^2\frac{q_0^2-|\vec{q}|^2}{2}\int\frac{d^3k}{(2\pi)^3}\frac{1}{E_{k}^+E^+_{k-q}}\theta(k_{F+}-|\vec{k}|)\delta(q_0-E^+_{k}-E^+_{k-q}) \nonumber\\
&-&  \pi G_1^2\frac{q_0^2-|\vec{q}|^2}{2}\int\frac{d^3k}{(2\pi)^3}\frac{1}{E^+_{k}E^+_{k-q}}\theta(k_{F+}-|\vec{k}|)\delta(q_0+E^+_{k}+E^+_{k-q}) \nonumber\\
&-&  \pi G_1^2\frac{q_0^2-|\vec{q}|^2}{2}\int\frac{d^3k}{(2\pi)^3}\frac{1}{E^+_{k}E^+_{k-q}}\theta(k_{F+}-|\vec{k}|)\delta(q_0-E^+_{k}+E^+_{k-q}) \nonumber\\
&+& \pi G_1^2\frac{q_0^2-|\vec{q}|^2}{2}\int\frac{d^3k}{(2\pi)^3}\frac{1}{E^+_{k}E^+_{k-q}}\theta(k_{F+}-|\vec{k}|)\delta(q_0+E^+_{k}-E^+_{k-q}) \nonumber\\
&+&\pi G_2^2\frac{q_0^2-|\vec{q}|^2-(m_++m_-)^2}{2}\int\frac{d^3k}{(2\pi)^3}\frac{1}{E_{k}^+E_{k-q}^-}\theta(k_{F+}-|\vec{k}|)\delta(q_0-E_{k}^+-E_{k-q}^-) \nonumber\\
&+&\pi G_2^2g^2\frac{q_0^2-|\vec{q}|^2-(m_++m_-)^2}{2}\int\frac{d^3k}{(2\pi)^3}\frac{1}{E_{k}^-E_{k-q}^+}\theta(k_{F-}-|\vec{k}|)\delta(q_0-E_{k}^--E_{k-q}^+) \nonumber\\
&-&  \pi G_2^2\frac{q_0^2-|\vec{q}|^2-(m_++m_-)^2}{2}\int\frac{d^3k}{(2\pi)^3}\frac{1}{E_{k-q}^+E_{k}^-}\theta(k_{F-}-|\vec{k}|)\delta(q_0+E_{k-q}^++E_{k}^-) \nonumber\\
&-&  \pi G_2^2\frac{q_0^2-|\vec{q}|^2-(m_++m_-)^2}{2}\int\frac{d^3k}{(2\pi)^3}\frac{1}{E_{k-q}^-E_{k}^+}\theta(k_{F+}-|\vec{k}|)\delta(q_0+E_{k-q}^-+E_{k}^+) \nonumber\\
&-&  \pi G_2^2\frac{q_0^2-|\vec{q}|^2-(m_++m_-)^2}{2}\int\frac{d^3k}{(2\pi)^3}\frac{1}{E_{k}^+E_{k-q}^-}\theta(k_{F+}-|\vec{k}|)\delta(q_0-E_{k}^++E_{k-q}^-) \nonumber\\
&-&  \pi G_2^2\frac{q_0^2-|\vec{q}|^2-(m_++m_-)^2}{2}\int\frac{d^3k}{(2\pi)^3}\frac{1}{E_{k}^-E_{k-q}^+}\theta(k_{F-}-|\vec{k}|)\delta(q_0-E_{k}^-+E_{k-q}^+) \nonumber\\
&+&  \pi G_2^2\frac{q_0^2-|\vec{q}|^2-(m_++m_-)^2}{2}\int\frac{d^3k}{(2\pi)^3}\frac{1}{E_{k-q}^+E_{k}^-}\theta(k_{F-}-|\vec{k}|)\delta(q_0-E_{k-q}^++E_{k}^-) \nonumber\\
&+&  \pi G_2^2\frac{q_0^2-|\vec{q}|^2-(m_++m_-)^2}{2}\int\frac{d^3k}{(2\pi)^3}\frac{1}{E_{k-q}^-E_{k}^+}\theta(k_{F+}-|\vec{k}|)\delta(q_0-E_{k-q}^-+E_{k}^+)  \nonumber\\
&+& \pi G_3^2\frac{q_0^2-|\vec{q}|^2}{2}\int\frac{d^3k}{(2\pi)^3}\frac{1}{E_{k}^-E^-_{k-q}}\theta(k_{F-}-|\vec{k}|)\delta(q_0-E^-_{k}-E^-_{k-q}) \nonumber\\
&-&  \pi G_3^2\frac{q_0^2-|\vec{q}|^2}{2}\int\frac{d^3k}{(2\pi)^3}\frac{1}{E^-_{k}E^-_{k-q}}\theta(k_{F-}-|\vec{k}|)\delta(q_0+E^-_{k}+E^-_{k-q}) \nonumber\\
&-&  \pi G_3^2\frac{q_0^2-|\vec{q}|^2}{2}\int\frac{d^3k}{(2\pi)^3}\frac{1}{E^-_{k}E^-_{k-q}}\theta(k_{F-}-|\vec{k}|)\delta(q_0-E^-_{k}+E^-_{k-q}) \nonumber\\
&+& \pi G_3^2\frac{q_0^2-|\vec{q}|^2}{2}\int\frac{d^3k}{(2\pi)^3}\frac{1}{E^-_{k}E^-_{k-q}}\theta(k_{F-}-|\vec{k}|)\delta(q_0+E^-_{k}-E^-_{k-q}) \ , \label{PiImaginary}
\end{eqnarray}
 where we have defined
 \begin{eqnarray}
 E_k^+ &\equiv& \sqrt{|\vec{k}|^2+m_+^2} \nonumber\\
 E_k^- &\equiv& \sqrt{|\vec{k}|^2+m_-^2}\ ,
 \end{eqnarray}
 and
 \begin{eqnarray}
A_+ &\equiv& \frac{q_0^2-|\vec{q}|^2+2|\vec{k}||\vec{q}|+2q_0E_k^+}{q_0^2-|\vec{q}|^2-2|\vec{k}||\vec{q}|+2q_0E_k^+}\frac{q_0^2-|\vec{q}|^2+2|\vec{k}||\vec{q}|-2q_0E_k^+}{q_0^2-|\vec{q}|^2-2|\vec{k}||\vec{q}|-2q_0E_k^+}\ , \\
A_{+-} &\equiv&\frac{q_0^2-|\vec{q}|^2+2|\vec{k}||\vec{q}|+m_+^2-m_-^2+2q_0E_k^+}{q_0^2-|\vec{q}|^2-2|\vec{k}||\vec{q}|+m_+^2-m_-^2+2q_0E_k^+}\frac{q_0^2-|\vec{q}|^2+2|\vec{k}||\vec{q}|+m_+^2-m_-^2-2q_0E_k^+}{q_0^2-|\vec{q}|^2-2|\vec{k}||\vec{q}|+m_+^2-m_-^2-2q_0E_k^+} \ ,\\
A_{-+} &\equiv&\frac{q_0^2-|\vec{q}|^2-2q_0E_k^-+2|\vec{k}||\vec{q}|+m_-^2-m_+^2}{q_0^2-|\vec{q}|^2-2q_0E_k^--2|\vec{k}||\vec{q}|+m_-^2-m_+^2}\frac{q_0^2-|\vec{q}|^2+2q_0E_k^-+2|\vec{k}||\vec{q}|+m_-^2-m_+^2}{q_0^2-|\vec{q}|^2+2q_0E_k^--2|\vec{k}||\vec{q}|+m_-^2-m_+^2} \ ,\\
A_{-} &\equiv&\frac{q_0^2-|\vec{q}|^2+2|\vec{k}||\vec{q}|+2q_0E^-_k}{q_0^2-|\vec{q}|^2-2|\vec{k}||\vec{q}|+2q_0E^-_k}\frac{q_0^2-|\vec{q}|^2+2|\vec{k}||\vec{q}|-2q_0E^-_k}{q_0^2-|\vec{q}|^2-2|\vec{k}||\vec{q}|-2q_0E^-_k}\ .
\end{eqnarray}
The couplings $G_1$, $G_2$ and $G_3$ are defined by
\begin{eqnarray}
G_1 &\equiv& g_{NN\pi}+4\sigma_0m_+h_{NN\pi} \ ,\\
G_2 &\equiv& g_{NN^*\pi}+2\sigma_0(m_--m_+)h_{NN^*\pi}\ , \\
G_3 &\equiv& g_{N^*N^*\pi}+4\sigma_0m_-h_{N^*N^*\pi}\ .
\end{eqnarray}

In a similar way, the retarded self-energy for $\eta$ meson in Fig.~\ref{fig:Eta} is
\begin{eqnarray}
{\rm Re}\tilde{\Sigma}^R_{\eta} (q_0,\vec{q})&=& \frac{g_{NN^*\eta}^2}{\pi^2}\int_0^{k_{F+}} d|\vec{k}| \frac{|\vec{k}|^2}{E_k^+}\left\{2-\frac{q_0^2-|\vec{q}|^2-(m_++m_-)^2}{4|\vec{k}||\vec{q}|}{\rm ln}\left|A_{+-}\right|\right\} \nonumber\\ \nonumber\\
&+& \frac{g_{NN^*\eta}^2}{\pi^2}\int_0^{k_{F-}} d|\vec{k}|\frac{|\vec{k}|^2}{E_k^-}\left\{2-\frac{q_0^2-|\vec{q}|^2-(m_++m_-)^2}{4|\vec{k}||\vec{q}|}{\rm ln}\left|A_{-+}\right|\right\}\ ,
\end{eqnarray}
and
\begin{eqnarray}
{\rm Im}\tilde{\Sigma}_{\eta}^R(q_0,\vec{q}) &=& \pi g_{NN^*\eta}^2\frac{q_0^2-|\vec{q}|^2-(m_++m_-)^2}{2}\int\frac{d^3k}{(2\pi)^3}\frac{1}{E_{k}^+E_{k-q}^-}\theta(k_{F+}-|\vec{k}|)\delta(q_0-E_{k}^+-E_{k-q}^-) \nonumber\\
&+&\pi g_{NN^*\eta}^2\frac{q_0^2-|\vec{q}|^2-(m_++m_-)^2}{2}\int\frac{d^3k}{(2\pi)^3}\frac{1}{E_{k}^-E_{k-q}^+}\theta(k_{F-}-|\vec{k}|)\delta(q_0-E_{k}^--E_{k-q}^+) \nonumber\\
&-&  \pi g_{NN^*\eta}^2\frac{q_0^2-|\vec{q}|^2-(m_++m_-)^2}{2}\int\frac{d^3k}{(2\pi)^3}\frac{1}{E_{k-q}^+E_{k}^-}\theta(k_{F-}-|\vec{k}|)\delta(q_0+E_{k-q}^++E_{k}^-) \nonumber\\
&-&  \pi g_{NN^*\eta}^2\frac{q_0^2-|\vec{q}|^2-(m_++m_-)^2}{2}\int\frac{d^3k}{(2\pi)^3}\frac{1}{E_{k-q}^-E_{k}^+}\theta(k_{F+}-|\vec{k}|)\delta(q_0+E_{k-q}^-+E_{k}^+) \nonumber\\
&-&  \pi g_{NN^*\eta}^2\frac{q_0^2-|\vec{q}|^2-(m_++m_-)^2}{2}\int\frac{d^3k}{(2\pi)^3}\frac{1}{E_{k}^+E_{k-q}^-}\theta(k_{F+}-|\vec{k}|)\delta(q_0-E_{k}^++E_{k-q}^-) \nonumber\\
&-&  \pi g_{NN^*\eta}^2\frac{q_0^2-|\vec{q}|^2-(m_++m_-)^2}{2}\int\frac{d^3k}{(2\pi)^3}\frac{1}{E_{k}^-E_{k-q}^+}\theta(k_{F-}-|\vec{k}|)\delta(q_0-E_{k}^-+E_{k-q}^+) \nonumber\\
&+&  \pi g_{NN^*\eta}^2\frac{q_0^2-|\vec{q}|^2-(m_++m_-)^2}{2}\int\frac{d^3k}{(2\pi)^3}\frac{1}{E_{k-q}^+E_{k}^-}\theta(k_{F-}-|\vec{k}|)\delta(q_0-E_{k-q}^++E_{k}^-) \nonumber\\
&+&  \pi g_{NN^*\eta}^2\frac{q_0^2-|\vec{q}|^2-(m_++m_-)^2}{2}\int\frac{d^3k}{(2\pi)^3}\frac{1}{E_{k-q}^-E_{k}^+}\theta(k_{F+}-|\vec{k}|)\delta(q_0-E_{k-q}^-+E_{k}^+) \ .
\end{eqnarray}
\end{widetext}
Thanks to a charge-conjugation invariance with respect to pion ($\eta$ meson), we have the following relations
\begin{eqnarray}
{\rm Re}\tilde{\Sigma}_{\pi(\eta)}^R(q_0,\vec{q}) &=& {\rm Re}\left(i\tilde{\Sigma}_{\pi(\eta)}(q_0,\vec{q})\right) \ , \\
{\rm Im}\tilde{\Sigma}_{\pi(\eta)}^R(q_0,\vec{q}) &=& \epsilon(q_0){\rm Im}\left(i\tilde{\Sigma}_{\pi(\eta)}(q_0,\vec{q})\right)\ .
\end{eqnarray}
By using these relations, we can get $\tilde{\Sigma}_{\pi(\eta)}(q_0,\vec{q})$. 

Note that in a limit of $q \to 0$, ${\rm Re}\tilde{\Sigma}^R_{\pi}(q_0,\vec{q})$ in Eq.~(\ref{PiReal}) vanishes together with the gap equation in Eq.~(\ref{GapEq}) in the chiral limit  $m_\pi\to0$. Namely, when we take infinite sums of self-energies in Fig.~\ref{fig:Pi}, pion becomes a massless particle which is consistent with a behavior of a Nambu-Goldstone (NG) boson.

\end{document}